\pdfoutput=1

\newcommand*{\ATLASLATEXPATH}{latex/}

\documentclass[cernpreprint, texlive=2016, UKenglish]{\ATLASLATEXPATH atlasdoc}
\usepackage{STDM-2017-09-PAPER-defs}
\usepackage{multirow}
\usepackage{\ATLASLATEXPATH atlaspackage}
\usepackage{\ATLASLATEXPATH atlasbiblatex}
\usepackage{\ATLASLATEXPATH atlasphysics}

\graphicspath{{logos/}{figures/}}

\usepackage{STDM-2017-09-PAPER-defs}

\AtlasTitle{Measurement of the four-lepton invariant mass spectrum in 13~TeV proton--proton collisions with the ATLAS detector}

\AtlasAbstract{%
A measurement of the four-lepton invariant mass spectrum is made with the ATLAS detector, using an integrated luminosity of 
36.1~fb$^{-1}$ of proton--proton collisions at $\sqrt{s}=13$~TeV delivered by the Large Hadron Collider.
The differential cross-section is measured for events containing two same-flavour opposite-sign lepton pairs. It exhibits a 
rich structure, with different mass regions dominated in the Standard Model by single $Z$ boson production, Higgs boson production, and $Z$ boson pair production, and non-negligible interference effects at high invariant masses.
The measurement is compared 
with state-of-the-art Standard Model calculations, which are found to be consistent with the data. 
These calculations are used to interpret the data in terms of $gg \rightarrow ZZ \rightarrow 4\ell$ and $Z \rightarrow 4\ell$ subprocesses, and to place constraints on a possible contribution from physics beyond the Standard Model.
}

\author{The ATLAS Collaboration}

\AtlasRefCode{STDM-2017-09}

\PreprintIdNumber{CERN-EP-2018-318}

 \AtlasJournal{JHEP}
\AtlasJournalRef{JHEP 04 (2019) 048}
\AtlasDOI{10.1007/JHEP04(2019)048}

\hypersetup{pdftitle={ATLAS document},pdfauthor={The ATLAS Collaboration}}

\begin{document}

\maketitle

\tableofcontents

\section{Introduction}
\label{sec:intro}
This paper presents a measurement of the four-lepton invariant mass ($\mFourL$) spectrum in events containing two same-flavour opposite-sign lepton (electron or muon) pairs.
The data correspond to 36.1~fb$^{-1}$ of proton--proton collisions collected with the ATLAS detector during the $\sqrt{s}=13~\TeV$ Large Hadron Collider (LHC) run in 2015--2016. 

\begin{figure}[hp]
        \centering
        \subfloat[]{\includegraphics[width = 0.25 \textwidth]{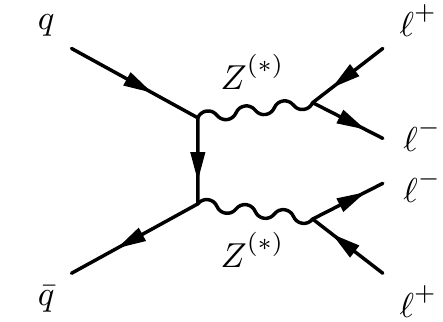}}
        \subfloat[]{\includegraphics[width = 0.25 \textwidth]{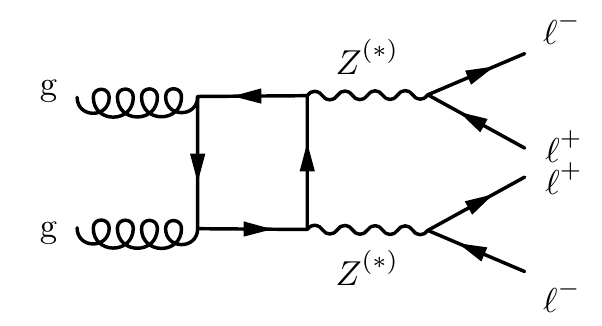}}
        \subfloat[]{\includegraphics[width = 0.25 \textwidth]{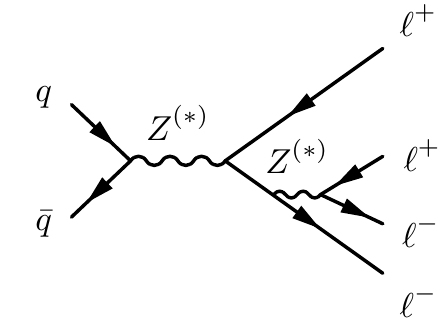}}
        \subfloat[]{\includegraphics[width = 0.25 \textwidth]{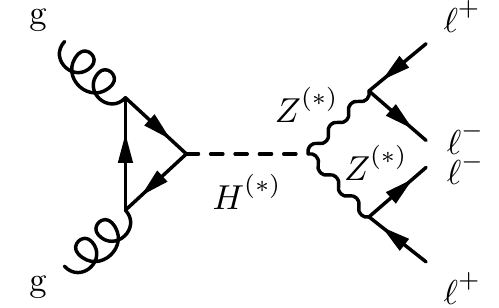}}

        \caption{Main contributions to the $pp \to 4\ell~(\ell = e,\mu)$ process: (a) $t$-channel \qqZZ production, (b) gluon-induced \ggZZ production via a quark loop, (c) internal conversion in $Z$ boson decays and (d) Higgs-boson-mediated $s$-channel production (here: gluon--gluon fusion). The notation $Z^{(\ast)}$ refers to a $Z$ boson which may be either on-shell or off-shell. \label{fig:ZZproduction} }

\end{figure}

\begin{figure}[h]
         \centering
         {\includegraphics[width = 0.66 \textwidth]{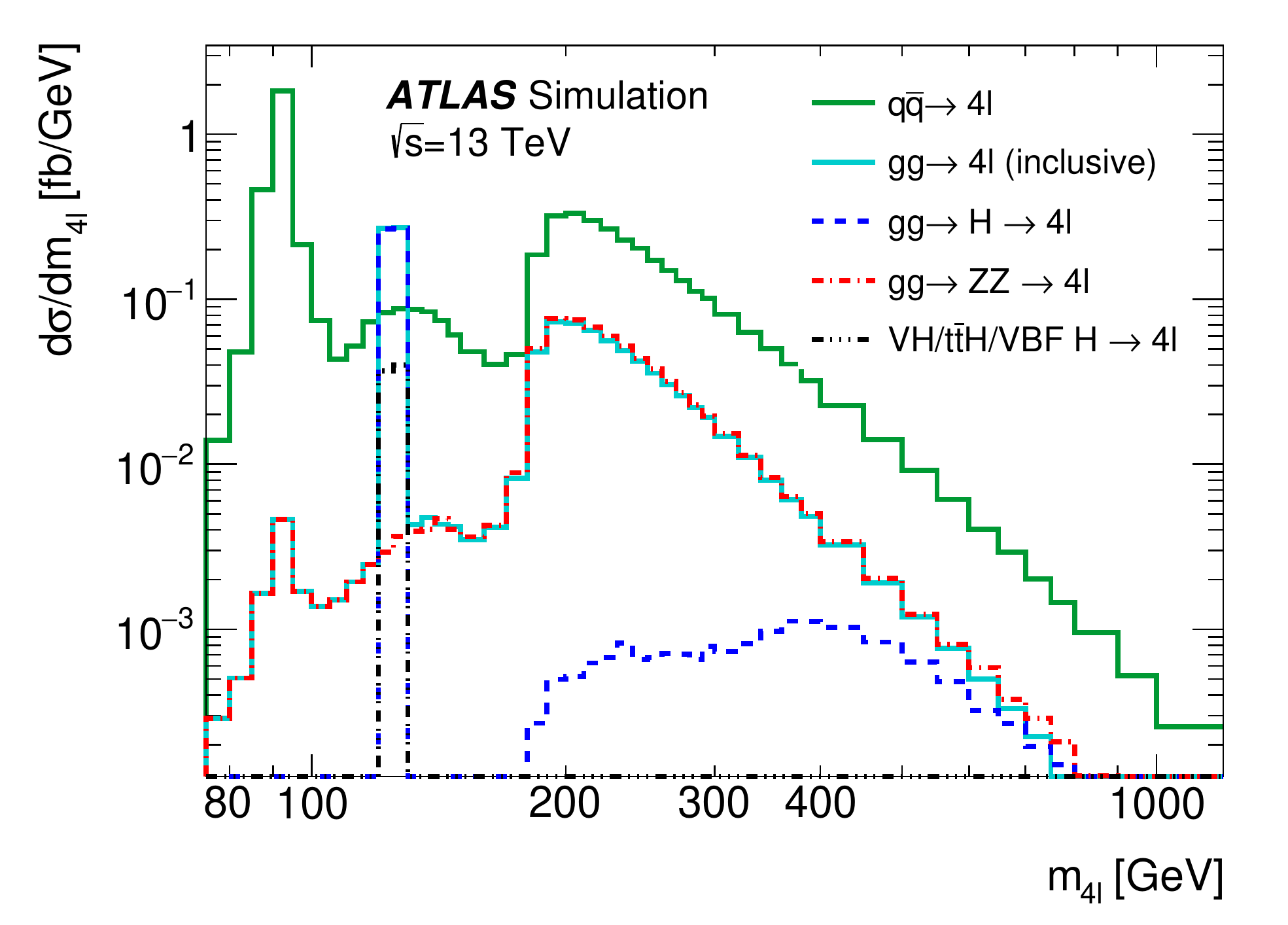}}
         \caption{Differential cross-sections as a function of the four-lepton invariant mass $\mFourL$ predicted by MC simulation. The total $gg\rightarrow 4\ell$ includes contributions from $gg\rightarrow H^{(\ast)} \rightarrow 4\ell$ as well as $gg \rightarrow 4\ell$ and the interference between the two. The $\qqZZ$ and $\ggZZ$ processes including off-shell Higgs boson production are modelled using $\SHERPAV{2.2.2}$ including all corrections described in Section~\ref{sec:MC}, while on-shell Higgs production is modelled using the dedicated samples based on $\POWHEG+\PYTHIAV{8}$ and \MGMCatNLO + \HERWIGpp described in the same section. \label{fig:lineshape} }
 \end{figure}

In $pp$ collisions four-lepton production is expected to receive contributions from several Standard Model (SM) physics processes, the most important of which are shown in  Figure~\ref{fig:ZZproduction}. The predicted cross-sections for these processes are shown as a function of the invariant four-lepton mass $\mFourL$ in Figure~\ref{fig:lineshape}.
Largest in magnitude is the quark-induced $t$-channel process \qqZZ, with leptonic ($\ell = e,\mu$) decays of the $Z$ bosons. 
Gluon-induced \ggZZ production also occurs, via an intermediate quark loop. The theoretical uncertainties in the SM prediction for 
this latter contribution are comparatively large. 

At around $\mFourL\simeq \mZ = 91.19$~\GeV~\cite{pdgreview}, single resonant \ZFourL production through QED radiative processes 
leads to a peak in the spectrum, and allows an extraction of the cross-section and branching fraction for $Z \rightarrow 4\ell$ to be made. 

Pairs of $Z$ bosons can also be produced from the decay of an intermediate Higgs boson. The majority of these are produced via gluon--gluon fusion, with minor contributions from
vector-boson fusion and associated production with vector bosons or top-quark pairs. 
There is resonant production around the Higgs boson mass of $m_{H} =124.97\pm0.24~\GeV$~\cite{HIGG-2016-33}, as well as off-shell production at higher mass values, 
which is enhanced at approximately 350~\GeV\ due to top-quark loops in the gluon--gluon fusion mechanism. 
At around 180~\GeV\ there is an enhancement of all the processes involving two $Z$ bosons, as on-shell production is possible above this mass. 

The box diagram $gg\rightarrow 4\ell$ and $gg\rightarrow H^{(*)}\rightarrow 4\ell$ processes interfere destructively in the SM. While interference is maximal around $\mFourL = 220~\GeV$~\cite{HIGG-2014-10}, the relative effect of the $gg\rightarrow H^{(*)}\rightarrow 4\ell$ contribution to the overall $\ggB$ lineshape is most pronounced above $350~\GeV$, as is visible in Figure~\ref{fig:lineshape}.

The off-shell Higgs production rate may be affected by beyond-the-SM (BSM) processes involving additional heavy particles, 
or modifications of the Higgs couplings, even if there is no effect on on-shell Higgs boson production~\cite{BSMhiggs}.

Previous measurements in this final state were carried out at $\sqrt{s}=13~\TeV$ by the ATLAS~\cite{STDM-2016-15} and CMS~\cite{CMS-SMP-16-017} collaborations with a focus on $ZZ$ production. The CMS result additionally includes a determination of the $Z\rightarrow 4\ell$ branching ratio using a dedicated detector-level analysis. 
The ATLAS Collaboration performed a measurement of inclusive four-lepton production at $\sqrt{s}=8~\TeV$~\cite{STDM-2014-15} and set constraints on the contribution from $\ggB$. An analysis using $\sqrt{s}=7~\TeV$ and $8~\TeV$ data~\cite{STDM-2013-08} to determine the $Z\rightarrow 4\ell$ branching fraction has also been published by ATLAS.  
Constraints on off-shell Higgs boson production have recently been set by ATLAS~\cite{HIGG-2017-06} using the $4\ell$ and $2\ell 2\nu$ final states in a dedicated detector-level analysis.  

This measurement is carried out in a fiducial phase space based on the kinematic acceptance of the detector to ensure a high selection efficiency.
The fiducial phase space and all observables are defined using stable final-state particles to minimise model dependence. 
The observation at detector level is corrected for experimental effects such as the detector and trigger system efficiencies and the detector resolution to provide results which may be used and reinterpreted without requiring a full simulation of the ATLAS detector. 
Electrons or muons originating from leptonic decays of the $\tau$-lepton are not considered to be part of the signal and their contribution to the observation at detector level is subtracted. 

Cross-sections are measured differentially in the invariant four-lepton mass $\mFourL$, and double-differentially with respect to 
both $\mFourL$ and the following kinematic variables: the transverse momentum of the four-lepton system $\ptFourL$, 
the rapidity of the four-lepton system $\yFourL$, and a matrix-element discriminant (introduced in Ref.~\cite{HIGG-2014-10}
and denoted by $D_\textrm{ME}$ in this paper) designed to distinguish
the $s$-channel Higgs-mediated production process from all other processes. 
The $\mFourL$ measurement is also made separately for each flavour combination of leptons in the event; $4e$, $4\mu$ and $2e2\mu$.
The double-differential cross-sections can provide additional sensitivity to the various subprocesses contributing to the measured 
final state; for example, the $\ptFourL$ is expected to discriminate $gg\rightarrow ZZ$ from $q\bar{q}\to ZZ$. They are also of interest for future interpretation; for example, some BSM contributions can have an impact which depends upon the final-state lepton flavours~\cite{Brooijmans:2018xbu}.
The measurements are compared with SM predictions. To explore the potential of reinterpreting differential cross-section measurements, they are also used to constrain the $gg \rightarrow 4\ell$ process and set a limit on the $gg \rightarrow H^{*} \rightarrow 4\ell$ off-shell signal strength, to extract the $Z \rightarrow 4\ell$ contribution and to place limits on a selected BSM scenario.
\FloatBarrier

\section{ATLAS detector}
\label{sec:detector}
\newcommand{\AtlasCoordFootnote}{%
ATLAS uses a right-handed coordinate system with its origin at the nominal interaction point (IP)
in the centre of the detector and the $z$-axis along the beam pipe.
The $x$-axis points from the IP to the centre of the LHC ring,
and the $y$-axis points upwards.
Cylindrical coordinates $(r,\phi)$ are used in the transverse plane, 
$\phi$ being the azimuthal angle around the $z$-axis.
The pseudorapidity is defined in terms of the polar angle $\theta$ as $\eta = -\ln \tan(\theta/2)$.
Angular distance is measured in units of $\Delta R \equiv \sqrt{(\Delta\eta)^{2} + (\Delta\phi)^{2}}$.}

The ATLAS experiment~\cite{PERF-2007-01,ATLAS-TDR-19,ATLAS-IBL} at the LHC is a multipurpose particle detector
with a forward--backward symmetric cylindrical geometry and a near $4\pi$ coverage in 
solid angle.\footnote{\AtlasCoordFootnote}
It consists of an inner tracking detector surrounded by a thin superconducting solenoid
providing a \SI{2}{\tesla} axial magnetic field, electromagnetic and hadron calorimeters, and a muon spectrometer.
The inner tracking detector covers the pseudorapidity range $|\eta| < 2.5$, and 
consists of silicon pixel, silicon microstrip, and transition radiation tracking detectors.
Lead/liquid-argon (LAr) sampling calorimeters provide electromagnetic (EM) energy measurements
with high granularity.
A hadron (steel/scintillator-tile) calorimeter covers the central pseudorapidity range ($|\eta| < 1.7$).
The endcap and forward regions are instrumented with LAr calorimeters
for both the EM and hadronic energy measurements up to $|\eta| = 4.9$.
The muon spectrometer (MS) surrounds the calorimeters and includes
three large air-core toroidal superconducting magnets with eight coils each.
The field integral of the toroids ranges between 2.0 and 6.0~Tm
across most of the detector. 
The MS is based on a system of precision chambers providing tracking information up to
$|\eta| = 2.7$ and fast detectors for triggering in the region $|\eta| < 2.4$.
A two-level trigger system is used to select events~\cite{TRIG-2016-01}.
The first-level trigger is implemented in hardware and processes a subset of the detector information
to reduce the accepted rate to at most \SI{100}{\kilo\hertz}.
This is followed by the software-based high-level trigger, which reduces the accepted event rate to \SI{1}{\kilo\hertz} on average
depending on the data-taking conditions.

\section{Definition of fiducial cross-section}
\label{sec:xsec}
The fiducial phase space used for the measurement is driven by the kinematic acceptance of the detector and closely follows the detector-level event selection described in Section~\ref{sec:eventsel}.
The kinematic selection is defined using stable final-state particles~\cite{ATL-PHYS-PUB-2015-013}.
Stable, prompt leptons (electrons and muons) are dressed by adding to their four-momenta the four-momenta of any photons not originating from hadron decays within a cone of size $\Delta R = \sqrt{(\Delta\eta)^{2} + (\Delta\phi)^{2}} = 0.005$ around the lepton direction. 
The fiducial phase space and any observables defined in this way are referred to as being at particle level. 
This definition is chosen to ensure that the particle-level distributions extrapolated from the detector-level observation are as model-independent as possible. This allows the extrapolation to be performed using detector resolutions and efficiencies which are known within experimentally controlled uncertainties, as described in Section~\ref{sec:unfolding}, without additional significant theoretical uncertainty.

Events are required to contain a quadruplet consisting of two same-flavour opposite-sign (SFOS) lepton pairs. 
The three leading leptons in the quadruplet must have transverse momenta ($\pt$) larger than 20, 15, and 10~\GeV, while the fourth lepton is required to have $\pt>7$~(5)~\GeV\ for electrons (muons). 
First, the lepton pair with an invariant mass closest to the $Z$ boson mass is selected as the primary dilepton pair with mass $m_{12}$. The remaining pair closest to the $Z$ boson mass is referred to as the secondary pair, with mass $m_{34}$, and completes the quadruplet. In this way, only one quadruplet is selected even in events containing more than four leptons. 
Requirements of $50 < m_{12} < 106$~\GeV\ and $f(\mFourL) < m_{34}< $115~\GeV\ are imposed, where the lower bound on $m_{34}$ is
calculated on an event-by-event basis as a function of the four-lepton invariant mass $\mFourL$, 
\begin{equation*}
        f(\mFourL) = \left\{ \begin{array}{ll} 5\text{ \GeV }, & \text{ for } \mFourL < 100\text{ \GeV }  \\  5\text{ \GeV } + 0.7 \times \left(\mFourL - 100\text{ \GeV }\right),  & \text{ for } 100\text{ \GeV } < \mFourL < 110\text{ \GeV } \\ 12\text{ \GeV }, & \text{ for } 110\text{ \GeV} <\mFourL < 140\text{ \GeV }  \\ 12\text{ \GeV } + 0.76 \times \left(\mFourL - 140\text{ \GeV\
     }\right),  & \text{ for } 140\text{ \GeV } < \mFourL < 190\text{ \GeV } \\ 50\text{ \GeV },  &\text{ for } \mFourL > 190\text{ \GeV } \\  \end{array} \right\}.
\end{equation*} 
This approach preserves high acceptance for low $\mFourL$ values, particularly for $Z\rightarrow 4\ell$, while suppressing events with leptons from leptonic $\tau$-lepton decays at higher values of $\mFourL$.

The angular separation between opposite flavour leptons in the quadruplet is required to satisfy $\Delta R > 0.2$, while any same flavour leptons have to be separated by $\Delta R > 0.1$ from each other. The latter condition enhances the acceptance for boosted topologies in high-$\mFourL$ $Z$ boson pair production. 
To exclude leptons originating from quarkonia decays, the invariant mass of any same-flavour, opposite-sign lepton pair in the event is required to exceed 5~\GeV.  A dedicated veto of leptons originating from $\Upsilon$ decays is not performed, in order to retain acceptance at low \mFourL, in particular for the single resonant $Z$ boson decay. This background is negligible within the phase space of this measurement.
The full list of selection criteria is given in Table~\ref{tab:fidsel} and largely follows Refs.~\cite{HIGG-2016-22,HIGG-2016-25}.
The overall range in $\mFourL$ considered for this measurement is $70~\GeV$ < \mFourL < 1200$~\GeV$ and was chosen based on the yields predicted in MC simulation. All candidates observed in the collision data fall into this interval. 

 \begin{table}[ht]
    \centering
    \caption{Definition of the fiducial region used for this measurement. All kinematic observables are defined using the dressed leptons. \label{tab:fidsel}}
    \begin{tabular}{ c  c}
    \hline \hline
         \multicolumn{2}{c}{\textbf \textsc{Physics Object Preselection}} \\
         \hline
         \hline 
        Muon selection &  $\pt > 5$~\GeV$, |\eta| < 2.7$  \\
        Electron selection &  $\pt > 7$~\GeV$, |\eta| < 2.47$  \\[0.3cm]
        \hline
        \hline 
        \multicolumn{2}{c}{\textbf \textsc{Quadruplet Selection}} \\
        \hline
        \hline 
        \multirow{3}{*}{Lepton pairing} & Assign SFOS lepton pairs with smallest \\
                                        & and second-smallest $|m_{\ell\ell}-m_{Z}|$ as \\
                                        & primary and secondary lepton pair, defining exactly one quadruplet \\ [0.3cm]
                                        \hline
        Lepton kinematics   &   $\pt > 20 / 15 / 10$~\GeV{} for leading three leptons \\ [0.3cm]
        Mass window, primary pair         &   $ 50$~\GeV$ < m_{12} < 106$~\GeV \\
        Mass window, secondary pair       &   $ f(\mFourL) < m_{34} < 115$~\GeV \\
        Lepton separation               &   $\Delta R_{ij} > 0.1 (0.2)$ for same (opposite) flavour leptons \\
        $J/\psi$ veto &    $  m_{ij} > 5$~\GeV~for all SFOS pairs \\
        Mass interval of measurement & 70~\GeV$ < \mFourL < 1200$~\GeV \\
    \hline\hline
    \end{tabular}
    \end{table}

    In addition to the invariant mass $\mFourL$, transverse momentum \ptFourL, rapidity $\yFourL$ and flavour composition of the selected quadruplet, the observables measured in this paper also include a matrix-element discriminant ($D_\textrm{ME}$) defined as 
    \begin{equation} \label{eq:MELAdeco1}
        D_\textrm{ME} = \log_{10} \frac{  \tilde{M}^{2}_{  gg\to H^{(\ast)} \to ZZ^{(\ast)} \to 4\ell} \left(p_{1,2,3,4}^{\mu}\right)}{  \tilde{M}^{2}_{ gg \left(\to H^{(\ast)} \right) \to ZZ^{(\ast)} \to 4\ell }\left(p_{1,2,3,4}^{\mu}\right) + 0.1 \cdot \tilde{M}^{2}_{q\bar{q}\to ZZ^{(\ast)} \to 4\ell }\left(p_{1,2,3,4}^{\mu}\right)  },
    \end{equation}
    with 
    \begin{equation*} \label{eq:MELAdeco2}
        \tilde{M}^{2}_{X} \left(p_{1,2,3,4}^{\mu}\right) =    \frac{\mathcal{M}^{2}_{ X} \left(p_{1,2,3,4}^{\mu}\right)  }{\left\langle\mathcal{M}^{2}_{X} \right\rangle\left(\mFourL \right)  },  
    \end{equation*}
    where $\mathcal{M}^{2}_{X} \left(p_{1,2,3,4}^{\mu}\right)$ indicates the squared matrix element for process $X$ evaluated for the specific four-momenta and flavours of the leptons in the given event, and $\left\langle\mathcal{M}^{2}_{X } \right\rangle\left(\mFourL \right) $ represents the average squared matrix element for process $X$ in the fiducial region for the given four-lepton invariant mass. 
    The first squared matrix element $\tilde{M}^{2}_{ gg \left(\to H^{(\ast)} \right) \to ZZ^{(\ast)}\to 4\ell}$ in the denominator of Eq.~(\ref{eq:MELAdeco1}) includes the non-Higgs box diagram (Figure~\ref{fig:ZZproduction}(b)), Higgs-mediated production (Figure~\ref{fig:ZZproduction}(d)), as well as the interference of the two, whereas the squared matrix element in the numerator $\tilde{M}^{2}_{ gg \to H^{(\ast)} \to ZZ^{(\ast)}\to 4\ell}$ only includes for Higgs-mediated production. 
    The constant factor multiplying the $t$-channel matrix element in the denominator affects the shape of the observable, but does not have a significant impact on its separation power. The value of 0.1 is chosen to keep the peak of the distribution sufficiently distant from the maximum possible value of 0 while also limiting tails in the negative direction. The numerator represents the $s$-channel matrix element involving the Higgs boson produced via gluon--gluon fusion. 
    The squared matrix elements are computed at leading-order QCD precision using the MCFM~\cite{Campbell:1999ah} program version 8.0. 
    The strong coupling constant is evaluated at the scale of half the four-lepton invariant mass. The Higgs boson mass is set to $m_{H} =  125.0~\GeV$, and its width to the Standard Model prediction for this mass. 
    Given the leading-order QCD precision, the incoming parton momenta are approximated by assuming the four-lepton centre-of-mass system is produced at rest.

\section{Data sample and event selection}
\label{sec:eventsel}
This measurement uses 36.1~fb$^{-1}$ of proton--proton collision data with a centre-of-mass energy $\sqrt{s}=13$~\TeV, collected during 2015 and 2016 with the ATLAS detector.

Events are selected in the online trigger system by requiring that one of several triggers be passed, in which 
one, two or three leptons (electrons or muons) are required with a range of lepton $\pt$ requirements dependent upon the multiplicity~\cite{HIGG-2016-19}.
The combined efficiency of these triggers for events within the detector-level phase space of the measurement is above $96\%$ for $70~\GeV < \mFourL < 180~\GeV$ and increases beyond 99\% for $\mFourL > 180~\GeV$ as the final-state leptons become more likely to satisfy the trigger thresholds. 

Electron identification is based on variables describing the longitudinal and transverse shapes of the electromagnetic showers in the calorimeters, properties of tracks in the inner detector, and track--cluster matching \cite{ATLAS-CONF-2014-032,ATL-PHYS-PUB-2015-041}.
Muons are identified using information from the muon spectrometer, the inner tracking detector and calorimeters, with the requirements depending upon the angular region and $\pt$ of the muon~\cite{PERF-2015-10}.

Using the candidates identified in this way, the detector-level event selection looks for four prompt leptons, as detailed in Table~\ref{tab:recosel}. 
Electrons are required to satisfy a loose-identification working point for which the efficiency is about 95\%~\cite{ATLAS-CONF-2016-024}, have $\ET > 7~\GeV{}$ and $|\eta|<2.47$.
Muons must likewise satisfy a loose-identification working point, designed to achieve high efficiencies of about $99\%$ with relatively low backgrounds~\cite{PERF-2015-10}, and have $\pt > 5~\GeV{}$, or $\pt > 15~\GeV{}$ if they are tagged solely in the calorimeter (``calorimeter-tagged muon''). 
To select leptons originating from the primary proton--proton interaction, their tracks are required to have a longitudinal impact parameter ($z_0$) satisfying $|z_{0}\sin(\theta)| < 0.5$~mm from the primary interaction vertex. 
Background from cosmic-ray muons is rejected by requiring each muon track's transverse impact parameter ($d_{0}$) to satisfy $|d_{0}|<1$~mm. This additionally discriminates against non-prompt muons. 

Using the leptons selected in this way, a quadruplet is formed according to the kinematic selection criteria defining the fiducial phase space described in Section~\ref{sec:xsec}. 
The quadruplet is then subjected to further requirements in order to suppress the contribution of leptons from secondary decays or misidentifications related to jet activity. 
It must not contain more than one muon identified solely in the calorimeter or solely in the muon spectrometer. 
None of the leptons constituting the quadruplet may have a transverse impact parameter significance $d_0/\sigma_{d_0} > 5~(3)$ for electrons (muons). 
All leptons of the quadruplet are required to satisfy isolation criteria based on particle-tracks measured in the inner detector and energy deposits in the electromagnetic calorimeter. When evaluating these criteria, tracks or deposits originating from leptons in the quadruplet are not considered in order to retain events with close-by prompt leptons.
Finally, the four leptons of the quadruplet are required to be loosely compatible with originating from a common vertex, evaluated by means of the reduced-$\chi^{2}$ vertex fit using the four lepton trajectories. This further suppresses the contribution of secondary leptons from $b$- and $c$-hadron decays.

\begin{table}[!htbp]
    \begin{center}
     \caption{Summary of the event selection requirements at detector level. \label{tab:recosel}}
    \vspace{0.2cm}
    \begin{tabular}{lll}
    \hline\hline
    \multicolumn{3}{c}{\textbf \textsc{Physics Object preselection}} \\
    \hline\hline
    & \multicolumn{1}{l}{\textsc{Electrons}} & \multicolumn{1}{l}{\textsc{Muons}} \\
    Identification & \textit{Loose} working point~\cite{ATLAS-CONF-2016-024} & \textit{Loose} working point~\cite{PERF-2015-10} \\  
    \multirow{2}{*}{Kinematics} &  \multirow{2}{*} {$\et > 7$~\GeV\ and $|\eta| < 2.47$} & 
             $\pt > 5$~\GeV\ and $|\eta| < 2.7$ \\
    &  &     $\pt > 15$~\GeV\ if calorimeter-tagged~\cite{PERF-2015-10} \\
    Interaction point constraint & $|z_{0} \cdot \sin\theta| < 0.5$~mm & $|z_{0} \cdot \sin\theta| < 0.5$~mm   \\
    Cosmic-ray muon veto & & $|d_0| < 1$~mm  \\
    \hline
    \hline 
    \multicolumn{3}{c}{\textbf \textsc{Quadruplet Selection}} \\
    \hline\hline 
    \textsc{Quadruplet formation}     & \multicolumn{2}{c}{Procedure and kinematic selection criteria as in Table~\ref{tab:fidsel}} \\
    \hline 
    \multicolumn{3}{l}{\textsc{Lepton isolation}} \\
    & \multicolumn{1}{l}{\textsc{Electrons}} & \multicolumn{1}{l}{\textsc{Muons}} \\
    Track isolation & $\sum\limits_{\Delta R\leq0.2} \pt < 0.15 \et^{e}$ & $\sum\limits_{\Delta R\leq0.3} \pt < 0.15\pt^{\mu}$ \\
    Calorimeter isolation & $\sum\limits_{\Delta R=0.2} \et < 0.2 \et^{e}$ & $\sum\limits_{\Delta R=0.2} \et <0.3 \pt^{\mu}$ \\
    & \multicolumn{2}{l}{\textit{Contributions from the other leptons of the quadruplet not considered}} \\
    \hline
    \multicolumn{3}{l}{\textsc{Lepton transverse impact parameter}} \\
    & \multicolumn{1}{l}{\textsc{Electrons}} & \multicolumn{1}{l}{\textsc{Muons}} \\
    & $d_0/\sigma_{d_0}<5$ & $d_0/\sigma_{d_0}<3$ \\ 
    \hline 
    \multicolumn{3}{l}{$4\ell$ \textsc{vertex fit}} \\
    $\chi^{2} / \mathrm{ndof}$ & \multicolumn{2}{c}{$ < 6$ ($4\mu$) or $< 9$ ($4e$, $2e2\mu$)} \\
    \hline\hline
    \end{tabular}
    \end{center}
    \end{table}

\section{Theoretical predictions and simulation}
\label{sec:MC}
Simulated events are used to correct the observed events for detector effects,
as well as to estimate the expected numbers of signal and background events and the systematic uncertainty of
the final results. Events from Monte Carlo simulation (MC) were passed through a detailed simulation of the ATLAS detector and trigger~\cite{SOFT-2010-01}, 
and the same reconstruction and analysis software as applied to the data. The effect of multiple $pp$ interactions per bunch crossing, as well as the effect on the detector response due to interactions from bunch crossings before or after the one containing the hard interaction, referred to as ``pile-up'', is emulated by overlaying inelastic $pp$ collisions onto the generated events. The events are then reweighted to reproduce the distribution of the number of collisions per bunch-crossing observed in the data. This procedure is known as ``pile-up reweighting''. To allow the contamination from events with $\tau$-leptons to be evaluated, generated samples include $\tau$-leptons.

The pair production of two $Z$ bosons via the $\qqZZ$ process was simulated with the \SHERPAV{2.2.2} event generator~\cite{Gleisberg:2008ta}.
Matrix elements were calculated for up to one parton at next-to-leading order (NLO) in QCD and up to three partons at
leading order (LO) using Comix~\cite{Gleisberg:2008fv} and OpenLoops~\cite{Cascioli:2011va}, and merged with the
\SHERPA\ parton shower~\cite{Schumann:2007mg} according to the ME+PS@NLO prescription~\cite{Hoeche:2012yf}.
The NNPDF3.0NNLO PDF set~\cite{Ball:2014uwa} was used, and the QCD renormalisation and factorisation scales were set to $\mFourL/2$.
The total cross-section from this calculation agrees within scale uncertainties with an NNLO QCD prediction obtained using the MATRIX program~\cite{Cascioli:2014yka,Grazzini:2015hta,Grazzini:2017mhc,Kallweit:2018nyv}.
A reweighting for virtual NLO EW effects~\cite{Biedermann:2016yvs,Biedermann:2016lvg} was applied as a function of the
four-lepton invariant mass, $\mFourL$, which modifies the differential cross-section by between $+3\%$ (for $\mFourL \sim 130~\GeV$) and $-20\%$ for $\mFourL>800~\GeV$.
The real higher-order electroweak contribution to $4\ell$ production in association with two jets (which includes vector-boson scattering) 
is not included in the sample discussed above but it was modelled separately using \SHERPAV{2.2.2} with the NNPDF3.0NNLO PDF set.
A second \qqZZ sample was generated at NLO precision in QCD using \POWHEGBOXV{v2}~\cite{Alioli:2010xd,Melia:2011tj,Nason:2013ydw} configured with the CT10 PDF set~\cite{Lai:2010vv} and interfaced to \PYTHIAV{8.186}\cite{Sjostrand:2006za,Sjostrand:2007gs} for parton showering.
A correction to higher-order precision ($K$-factor), defined for this process as the ratio of the cross-section at NNLO QCD accuracy to the one at NLO QCD accuracy, was obtained using the MATRIX NNLO QCD prediction and applied to this sample as a function of $\mFourL$, modifying the inclusive cross-section by between $+10\%$ for $\mFourL < 180~\GeV$ and $+25\%$ for $\mFourL >~800~\GeV$. The reweighting for virtual NLO EW effects discussed above for the \SHERPA case was also applied to this sample.

The purely gluon-initiated $ZZ$ production process enters at next-to-next-to-leading order (NNLO) in $\alphas$. It was modelled using
\SHERPAV{2.2.2}~\cite{Cascioli:2013gfa}, at LO precision for zero- and one-jet final states, and the NNPDF3.0NNLO PDF set was chosen.
This sample includes the box diagram, the $s$-channel process proceeding via a Higgs boson, and the interference between the two. 
Recently, a NLO QCD calculation for the three components became 
available~\cite{Caola2015,Caola2016} allowing $\mFourL$ differential $K$-factors to be
calculated with the 1/$m_{t}$ expansion below 2$m_{t}$, and assuming a massless quark approximation
above this threshold. This NLO QCD calculation was used to correct the $s$-channel process \ggS{}, the box diagram \ggB{} and the interference with separate $K$-factors. These represent significant corrections of the order of $+100\%$ to the leading-order cross-section.
There are, however, NNLO QCD precision calculations for the off-shell Higgs boson production cross-section~\cite{Passarino:2013bha,YR4} which show additional enhancement of the cross-section.
Since these corrections are not
known differentially in $\mFourL$ for all three components, the prediction for each component is scaled by an additional overall correction factor of 1.2, assumed to be the same for the signal, background and interference.
This additional constant
scale factor is justified by the approximately constant behaviour of the NNLO/NLO QCD prediction. In addition, a purely leading-order prediction for the $\ggB$ process was obtained using the MCFM program~\cite{Campbell:1999ah} with the CT10 PDF set~\cite{Lai:2010vv}, interfaced to \PYTHIAV{8}\cite{Sjostrand:2006za,Sjostrand:2007gs}.

In the mass range $100~\GeV < \mFourL < 150~\GeV$, where on-shell Higgs production dominates and the effect of interference is negligible, dedicated samples are used to model the on-shell Higgs and box diagram continuum $ZZ$ production processes. 
In the case of the box diagram, the same combination of NLO QCD $K$-factor and a factor of 1.2 to account for higher-order effects, as described above, is applied to correct the cross-section. 
The Higgs production processes via gluon--gluon fusion (ggF)~\cite{Campbell:2012am} (which dominates the on-shell Higgs production), via vector-boson
fusion (VBF)~\cite{Nason:2009ai} and in association with a vector boson ($VH$)~\cite{Luisoni:2013kna}
were all simulated at NLO precision in QCD using \POWHEGBOXV{v2} with the PDF4LHC next-to-leading-order (NLO) set of parton distribution functions\,\cite{Butterworth:2015oua} and interfaced to \PYTHIAV{8.186}. 
The decay of the Higgs and $Z$ bosons was performed within \PYTHIA. 
The description of the gluon--gluon fusion process was further improved by reweighting to NNLO QCD accuracy using the HNNLO program~\cite{Catani:2007vq,Grazzini:2008tf,Grazzini:2013mca}, referred to as the NNLOPS method~\cite{Hamilton:2013fea}, and the resulting prediction was normalised using cross-sections calculated at N3LO precision in QCD~\cite{YR4}.
For VBF production, full NLO QCD and EW calculations were used with approximate NNLO QCD corrections. The VH production was calculated at NNLO in QCD and NLO EW corrections are applied. 
Production in association with a top-quark pair was
simulated to NLO accuracy in QCD using \MGMCatNLO~\cite{Wiesemann:2014ioa,Alwall:2014hca} configured with the CT10 PDF set and interfaced
to \HERWIGpp~\cite{Bahr:2008pv,Bellm:2013hwb}. The contribution from this process is very small in the analysis.

Other SM processes resulting in four prompt leptons in the final state are considered as irreducible backgrounds, and were also simulated
using MC generators. These include triboson production ($ZWW$, $ZZW$ and $ZZZ$) and $t\bar{t}$ pairs produced in association with vector bosons ($t\bar{t}Z$,
$t\bar{t}WW$) collectively referred to as $t\bar{t}V(V)$. The triboson processes were generated with \SHERPAV{2.1.1} using the CT10 PDF set.
The $WWZ$ prediction has leading-order QCD precision for up to two additional outgoing partons while the $WZZ$ and $ZZZ$ prediction has next-to-leading-order QCD precision for zero additional outgoing partons and leading-order QCD precision for up to two partons. The $t\bar{t}V$ processes were generated with \SHERPAV{2.2.0} at leading-order QCD precision and the NNPDF3.0NNLO PDF set.

In addition to these contributions, reducible background processes which can contribute to the final event selection but contain 
at least one non-prompt or mis-reconstructed lepton are estimated using a partially data-driven method detailed in Refs.~\cite{HIGG-2016-22,HIGG-2016-25}. These processes include one or more leptons produced from heavy-flavour 
hadron decays, muons from pion or kaon decays, or electrons from either photon conversion or hadron misidentification. 
The majority of these events originate from $Z$ bosons produced in association with jets, $t\bar{t}$ production with leptons from heavy-flavour
decay, and $WZ$ production in association with jets. Contributions from these processes are estimated separately depending on the flavour of the leptons in the secondary pair and
the source of the non-prompt lepton(s). This estimation procedure uses a number of different control regions and simultaneous fits, and for some specific processes the
estimation is taken directly from MC simulation. The data-driven results were validated in separate control regions using data. This contribution is small compared to that of prompt four-lepton production, and negligible for $\mFourL > 200~\GeV$.

\section{Unfolding for detector effects}
\label{sec:unfolding}
The measured four-lepton mass spectrum and additional double-differential spectra are ``unfolded'' to correct for experimental effects, 
including the resolution and efficiency of the detector and trigger system. This allows direct comparison with particle-level predictions within
the fiducial phase space.

The unfolding procedure is based on describing the relationship between the number of events measured in a bin $d$ of a particular detector-level differential distribution and the yield in bin $p$ of the corresponding particle-level distribution using a single response matrix $R_{dp}$.
This matrix consists of three contributions: 
\begin{itemize}
\item The reconstruction efficiency is measured as the ratio of the number of events which pass both the fiducial and detector event selections to the number passing the fiducial selection, as a function of the kinematic observable(s) at particle level. Above $\mFourL = 200$~\GeV, it is typically between $60\%$ and $80\%$, while for lower values of $\mFourL$, values as low as $30\%$ are reached for the $4e$ final state, due to reduced detector efficiency when reconstructing leptons of low transverse momenta. It enters $R_{dp}$ as a diagonal matrix.

\item A ``migration matrix'' which contains the probabilities that a particle-level event from a given fiducial bin which passes the detector selection will be found in a particular reconstructed bin. It accounts for bin-to-bin migrations. For all measurements, the diagonal elements of this matrix, also referred to as the ``fiducial purity'' in each bin, have values above $80\%$, with most of the small amount of migration occurring between neighbouring mass bins.  
 
\item Finally, the fiducial fraction accounts for events which pass the detector selection but fail the fiducial event selection. 
This can occur due to the resolution of the detector, or leptons originating from leptonically decaying $\tau$-leptons. 
It is measured by taking the ratio of events which pass both the fiducial and detector selection to the total passing the detector selection. It is close to unity for $\mFourL > 200$~\GeV, and above $90\%$ below this threshold. It enters $R_{dp}$ as a diagonal matrix.

\end{itemize}

In the unfolding procedure, first, the fiducial fraction is accounted for by multiplying the background-subtracted observation in each bin of the measurement with the fiducial fraction for that particular bin. 
Then, an iterative Bayesian procedure~\cite{DAgostini:1994fjx}, using the particle-level predicted distribution as the initial prior and the migration matrix, is used to correct for bin migration. The iteration procedure reduces the dependence on the initial prior. The number of iterations is used as a regularisation parameter and controls the statistical uncertainty. Two iterations are found to be optimal for all distributions by MC studies aiming to minimise both the statistical uncertainty and the bias. 
Finally, the resulting estimate of the particle-level distribution is divided by the reconstruction efficiency bin by bin to obtain the final result. 
This approach represents a compromise between accounting for the small migration effects that occur and minimising the effect of small fluctuations in the detector-level distributions through the regularisation approach.   

The binning used for the measurements presented in this paper is driven by the requirements of the procedure described above. Bin edges are placed to cover as wide as possible a phase-space interval with fine granularity while ensuring a fiducial purity of at least $80\%$. In addition, a minimum predicted detector-level yield of 10 events is required in each bin to ensure the numerical stability of the unfolding procedure and the viability for reinterpretation.

The robustness of the unfolding procedure to possible deviations of the data from the SM prediction was studied to ensure the model-independence of the analysis. Three scenarios were checked by unfolding pseudo-data after including the following: a greatly varied rate from off-shell Higgs production, or gluon-induced $ZZ$ production, ($-75\%$ / $+200\%$ and $-100\%$ / $+400\%$ respectively) and the injection of an additional scalar resonance (masses of 200, 400 and 900~\GeV\ were used). For the smooth, non-resonant modifications of the lineshape, the true lineshape was reproduced by unfolding with the SM-based response matrix with excellent accuracy, with residual biases far less than statistical precision. For large, resonant BSM contributions the bias is larger, up to the order of the statistical uncertainty when using the high-$D_\textrm{ME}$ region (defined in Section~\ref{sec:results}). This type of interpretation is not considered here, but it is noted for any reinterpretations which may be affected.

\section{Uncertainties}
\label{sec:uncertainties}
The limiting source of uncertainty in this measurement is the statistical uncertainty, which is many times greater than the total systematic uncertainty in some bins. Experimental and theoretical sources both contribute to the systematic uncertainty, and their relative impact varies depending on the bin.

The statistical uncertainty of the data is estimated using 2000 Poisson-distributed pseudo-datasets centred on the observed value in each bin, and repeating the unfolding procedure for each set.
The root mean square of the differences between the resulting unfolded distributions and the unfolded data is taken as the statistical uncertainty in each bin.

Experimental systematic uncertainties affect the response matrix used in the unfolding procedure.
They are dominated by the reconstruction, identification and isolation efficiency uncertainties for electrons~\cite{ATLAS-CONF-2016-024, PERF-2013-05} and muons~\cite{PERF-2015-10}. There are smaller contributions from lepton momentum resolution and scale uncertainties, and the uncertainty in the pile-up reweighting.

The uncertainty in the combined 2015+2016 integrated luminosity is 2.1$\%$. It is derived, following a methodology similar to that detailed in Ref.~\cite{DAPR-2013-01}, and using the LUCID-2 detector for the baseline luminosity measurements \cite{LUCID2}, from calibration of the luminosity scale using x-y beam-separation scans. This uncertainty is fully correlated across all measured cross-section bins and is propagated to the limit setting in the interpretations of the results.
All other sources of systematic uncertainty are propagated to the final unfolded
distributions by varying the inputs within their uncertainty, repeating the unfolding, and taking in each bin
the resulting deviation from the nominal response matrix as the uncertainty.

Theoretical uncertainties primarily affect the particle-level predictions obtained from simulation. Since they affect the contribution of individual subprocesses to the total cross-section and the final-state lepton kinematics, they also impact the response matrix and hence the measured cross-sections.
However, this is a very small effect compared to the experimental uncertainties and the statistical uncertainty. The most significant sources of theoretical uncertainty are the choice of factorisation and renormalisation scales, PDF set, and parton showering model within the event generator for the \qqZZ and \ggZZ MC samples.

In the case of \qqZZ , the full uncertainty due to the scale choice was estimated using seven sets of values for the renormalisation and factorisation scales obtained by independently varying each to either one half, one, or two times the nominal value while keeping their ratio in the range of $[0.5,2]$.
Since a NLO QCD $K$-factor obtained within the fiducial phase space is applied in the \ggZZ samples, the uncertainty due to the scale choice for this production process within the fiducial phase space is evaluated using the differential scale uncertainty of this $K$-factor. In addition, seven sets of two values for the scales as described above are used to evaluate the impact of the scale choice on the acceptance for \ggZZ.

Due to the reweighting of the purely gluon-induced $ZZ$ production processes described in Section~\ref{sec:MC}, there are several other uncertainties affecting the normalisation in addition to the scale-induced uncertainties calculated together with the NLO QCD $K$-factors discussed above. In the $\mFourL$ region below $2 m_{t}$, the higher-order corrections were computed solely for events not featuring jets with $\pt > 150~\GeV$ to ensure a good description by the $1/m_{t}$ expansion. Therefore, the default scale uncertainty is doubled for about $8\%$ of the events in this region which contain such jets. Likewise, the scale uncertainty is also doubled at $2 m_{t}$, with a Gaussian-smoothed transition from this maximal value down to the default uncertainty within a distance of $50$~\GeV\ to either side of the threshold. The inflated uncertainty is intended to account for potential effects as the top quarks become on-shell. It is assumed that the relative NLO QCD corrections for massless and massive loops behave similarly beyond $2 m_{t}$ and that the NNLO QCD correction calculated for the off-shell Higgs production process mimics the continuum production and the interference well, so no further uncertainty is considered. It is expected that the NLO QCD scale uncertainty covers these effects, as it is larger than the one calculated at NNLO QCD.

The uncertainty due to the choice of PDF set was estimated for both \qqZZ{} and \ggZZ{} by reweighting the sample to the alternative PDF sets CT10 and MSTW~\cite{Martin:2009iq} as well as evaluating eigenvector variations of the default NNPDF3.0NNLO PDF set.
In the case of \qqZZ{}, the envelope of these three variations is used to assign an uncertainty.
For \ggZZ{}, the envelope is formed using only the effect of the variations on the shapes, as the cross-section is taken from the higher-order reweighting.

The impact on the detector corrections originating from differences in the showering model was assessed for both processes by varying the CKKW matching scale~\cite{Catani:2001cc,Krauss:2002up} from the \SHERPAV{2.2.2} default, changing the dipole recoil scheme in the shower to the one in \cite{Hoeche:2010} and by varying the resummation scale up and down by a factor of two. Furthermore, in order to account for non-factorising effects, \qqZZ{} events with high QCD activity~\cite{Gieseke:2014gka}
were assigned an additional uncertainty of the size of the NLO EW correction.
As the NLO EW reweighting is only applied for \qqZZ , this last uncertainty is
not applied to the \ggZZ{} or $gg\rightarrow H^{(\ast)} \rightarrow 4\ell$ processes.

Theoretical uncertainties in the modelling of resonant Higgs boson production do not have a significant effect on the response matrix, since this process is confined to a single bin in the $\mFourL$ spectrum. They mainly affect the predicted particle-level differential cross-sections. The same uncertainties as reported in Ref.~\cite{HIGG-2016-22} are applied in this paper. They are dominated by QCD scale and PDF uncertainties affecting the gluon--gluon fusion component.

In order to cross-check and estimate the uncertainty due to the choice of generator used to model the \qqZZ process, the difference
between the unfolded results using the nominal \SHERPAV{2.2.2} samples and the alternative \POWHEG + \PYTHIAV{8} sample
is taken as a systematic uncertainty. 

The MC statistical uncertainty in the unfolding procedure is evaluated using a bootstrap method with 2000 toy samples, each assigning a Poisson weight with an expected value of one to every MC event used in the analysis. The RMS of the unfolded result in each bin for all toy samples is then taken as an uncertainty, and is typically between 0.5$\%$ and 1.5$\%$ per bin.

The uncertainty due to the unfolding method itself is estimated as follows. The MC events are reweighted with
fitted functions of the particle-level observables to give good agreement between the reconstructed MC distribution and the observed data distribution.
The reconstructed MC distribution is then unfolded using the nominal response matrix and compared with the reweighted particle-level distribution,
with the difference between the two taken as a systematic uncertainty in each bin.
For the majority of bins this is less than 1$\%$, with the exception of two bins with the fewest number of events in the double-differential $\mFourL$--$\ptFourL$ distribution (defined in Section~\ref{sec:results}) which result in 3$\%$ and 5$\%$ uncertainties. For comparison, the statistical uncertainty is around 25$\%$ and 45$\%$ in those respective bins.

\begin{figure}[h]
    \centering
        \subfloat{\includegraphics[width = 0.72 \textwidth]{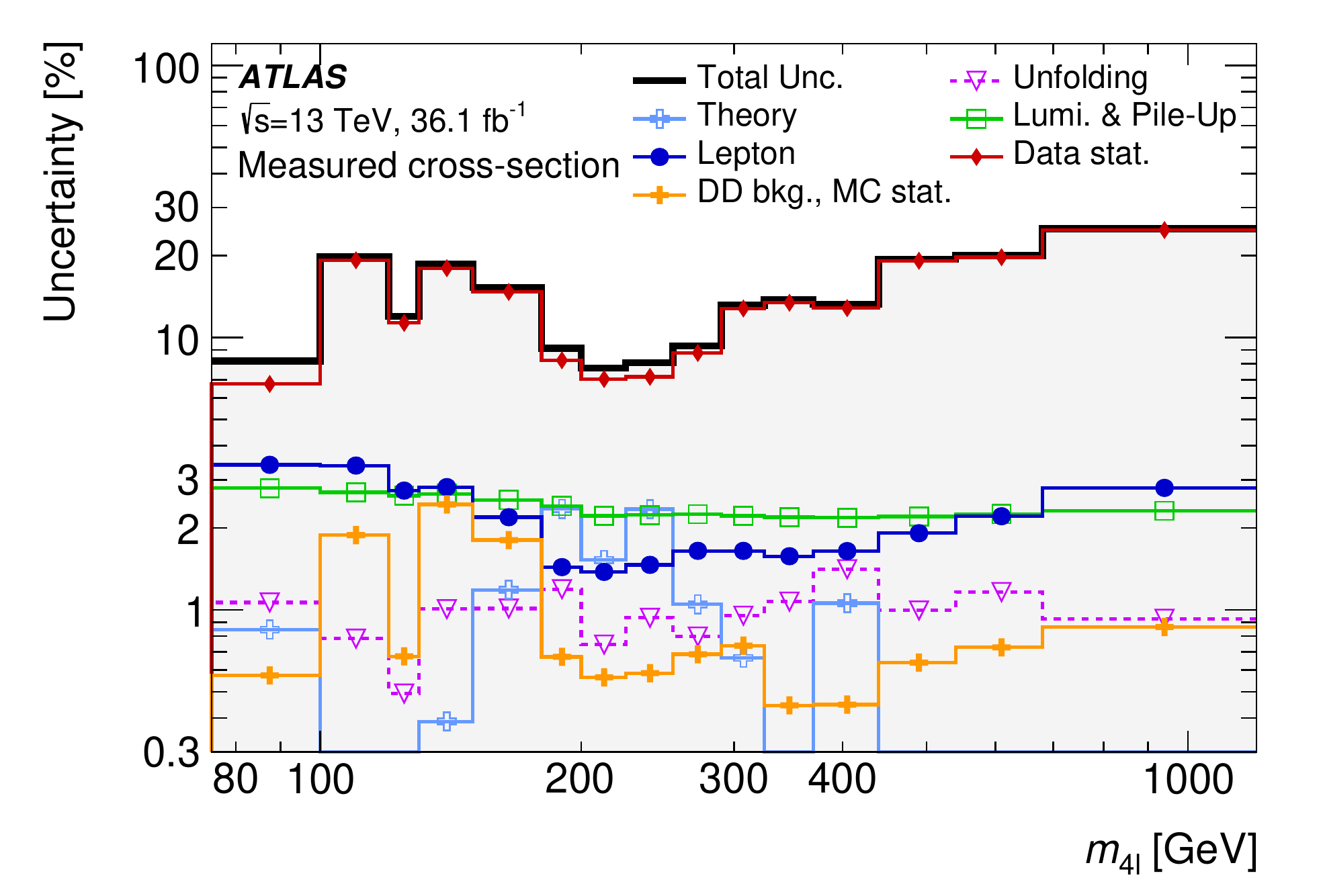} }
\caption{The leading sources of uncertainty in the measured cross-section after unfolding are given in percent as a function of the four-lepton invariant mass. The ``Unfolding'' category includes the effect of the generator choice for $\qqZZ$ and the uncertainty due to the unfolding method itself, added in quadrature.
         The ``Lepton'' category comprises the lepton reconstruction and selection efficiencies as well as momentum resolution and scale uncertainties. ``DD bkg'' refers to the data-driven estimation used for the reducible background contribution.
            \label{fig:sysm4l}}
 \end{figure}

\begin{figure}[h]
    \centering
        \subfloat[double-differential $\ptFourL$-$\mFourL$]{\includegraphics[width = 0.49 \textwidth]{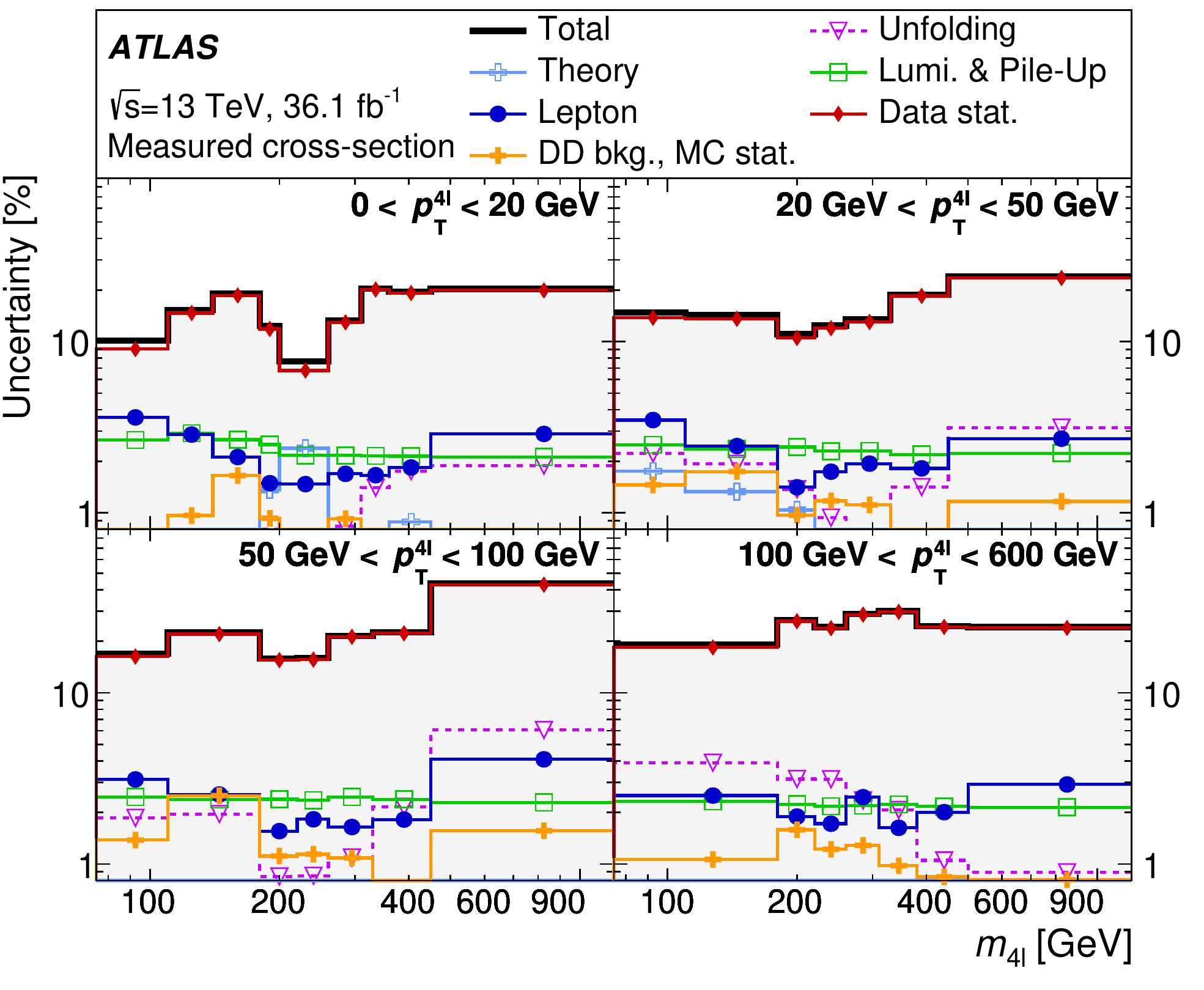} }
        \subfloat[double-differential $|\yFourL|$-$\mFourL$]{\includegraphics[width = 0.49 \textwidth]{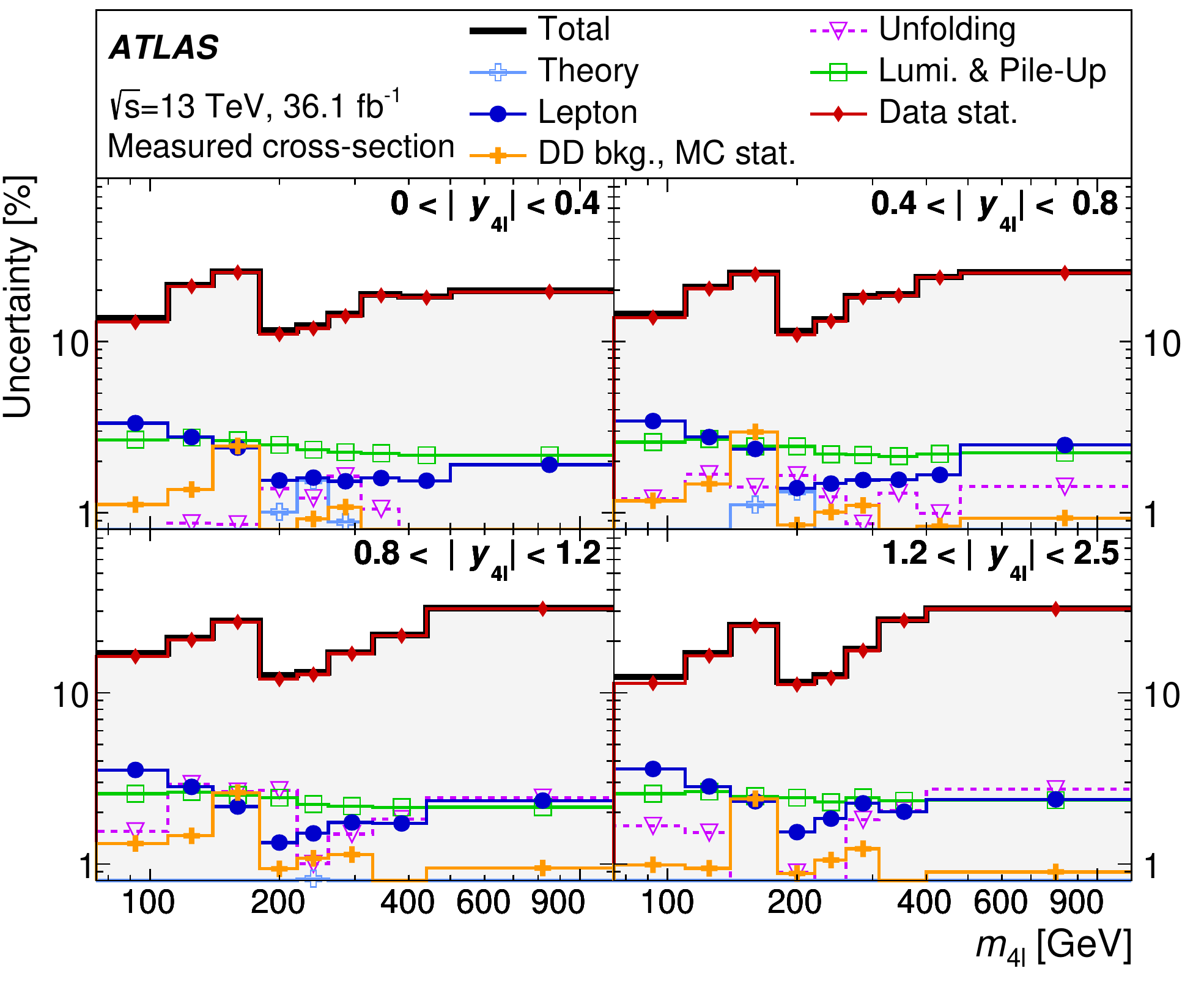} }
\caption{
        The leading sources of uncertainty in the measured cross-section after unfolding are given in percent as a function of (a) the four-lepton invariant mass in slices of $\ptFourL$ and (b) the four-lepton invariant mass in slices of $|\yFourL|$. The ``Unfolding'' category includes the effect of the generator choice for $\qqZZ$ and the uncertainty due to the unfolding method itself, added in quadrature.
         The ``Lepton'' category comprises the lepton reconstruction and selection efficiencies as well as momentum resolution and scale uncertainties. ``DD bkg'' refers to the data-driven estimation used for the reducible background contribution.
            \label{fig:sysm4l_2}}
 \end{figure}

\begin{figure}[h]
    \centering
        \subfloat[double differential $D_\textrm{ME}$-$\mFourL$]{\includegraphics[width = 0.49 \textwidth]{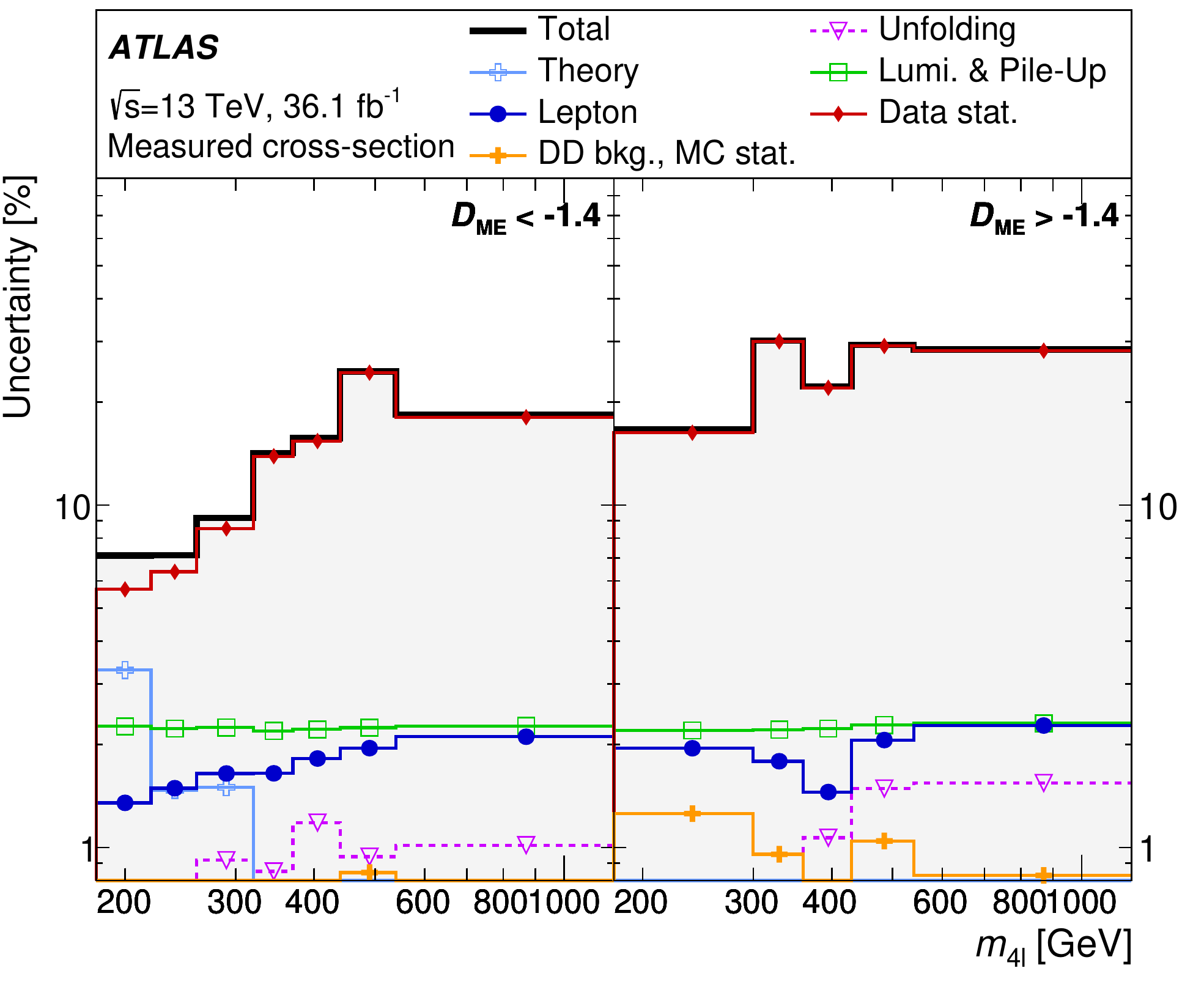} }
        \subfloat[$\mFourL$ per lepton flavour channel]{\includegraphics[width = 0.49 \textwidth]{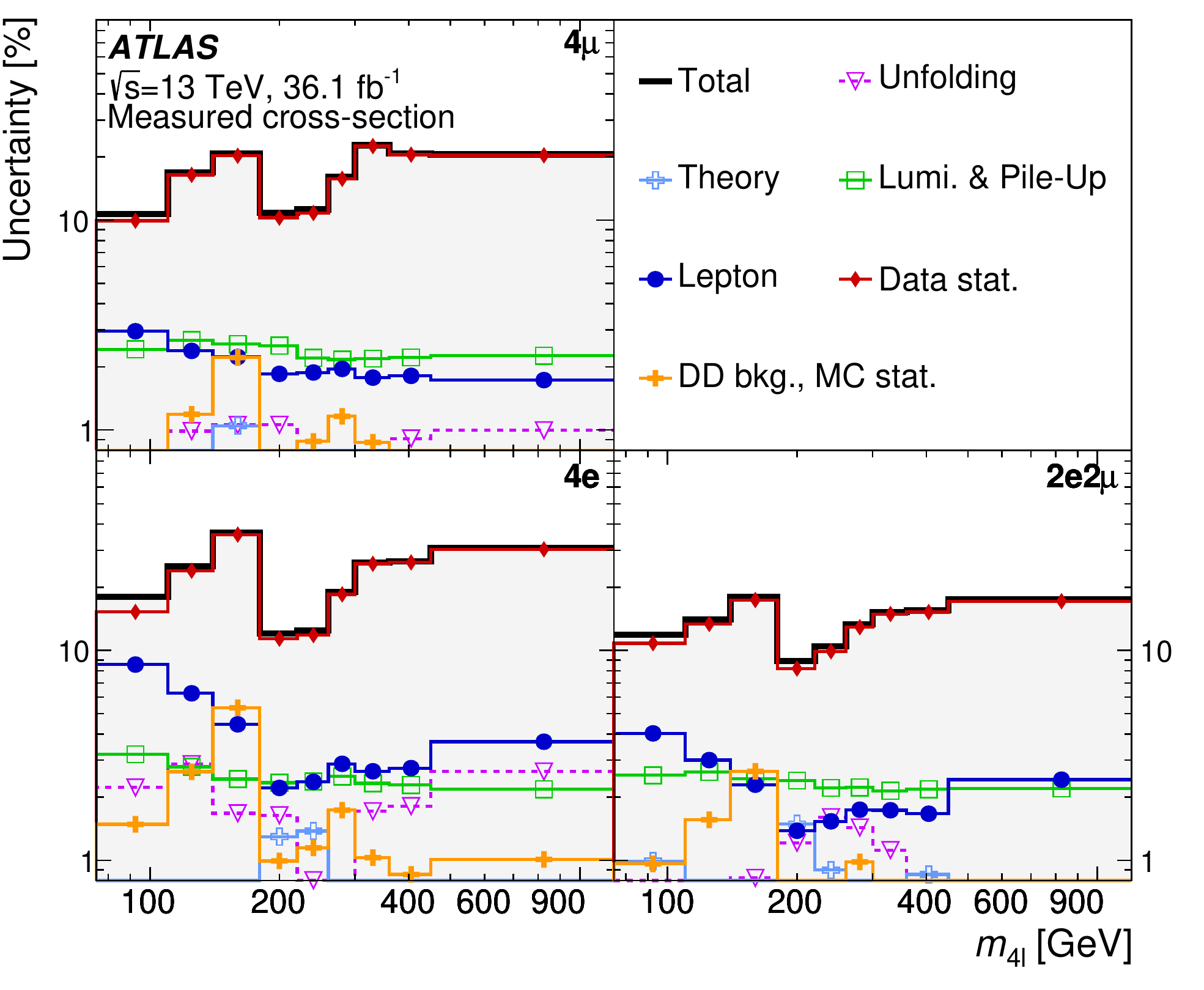} }
\caption{
        The leading sources of uncertainty in the measured cross-section after unfolding are given in percent as a function of (a) the four-lepton invariant mass in slices of the $D_\textrm{ME}$ discriminant and (b) the four-lepton invariant mass per final-state flavour channel. The ``Unfolding'' category includes the effect of the generator choice for $\qqZZ$ and the uncertainty due to the unfolding method itself, added in quadrature.
         The ``Lepton'' category comprises the lepton reconstruction and selection efficiencies as well as momentum resolution and scale uncertainties. ``DD bkg'' refers to the data-driven estimation used for the reducible background contribution.
            \label{fig:sysm4l_3}}
 \end{figure}

The various contributions to the uncertainties in the final result are summarised in Figures~\ref{fig:sysm4l}--\ref{fig:sysm4l_3}.

\section{Measured distributions}
\label{sec:results}
\begin{figure}[h]
    \centering
        {\includegraphics[width = 0.66 \textwidth]{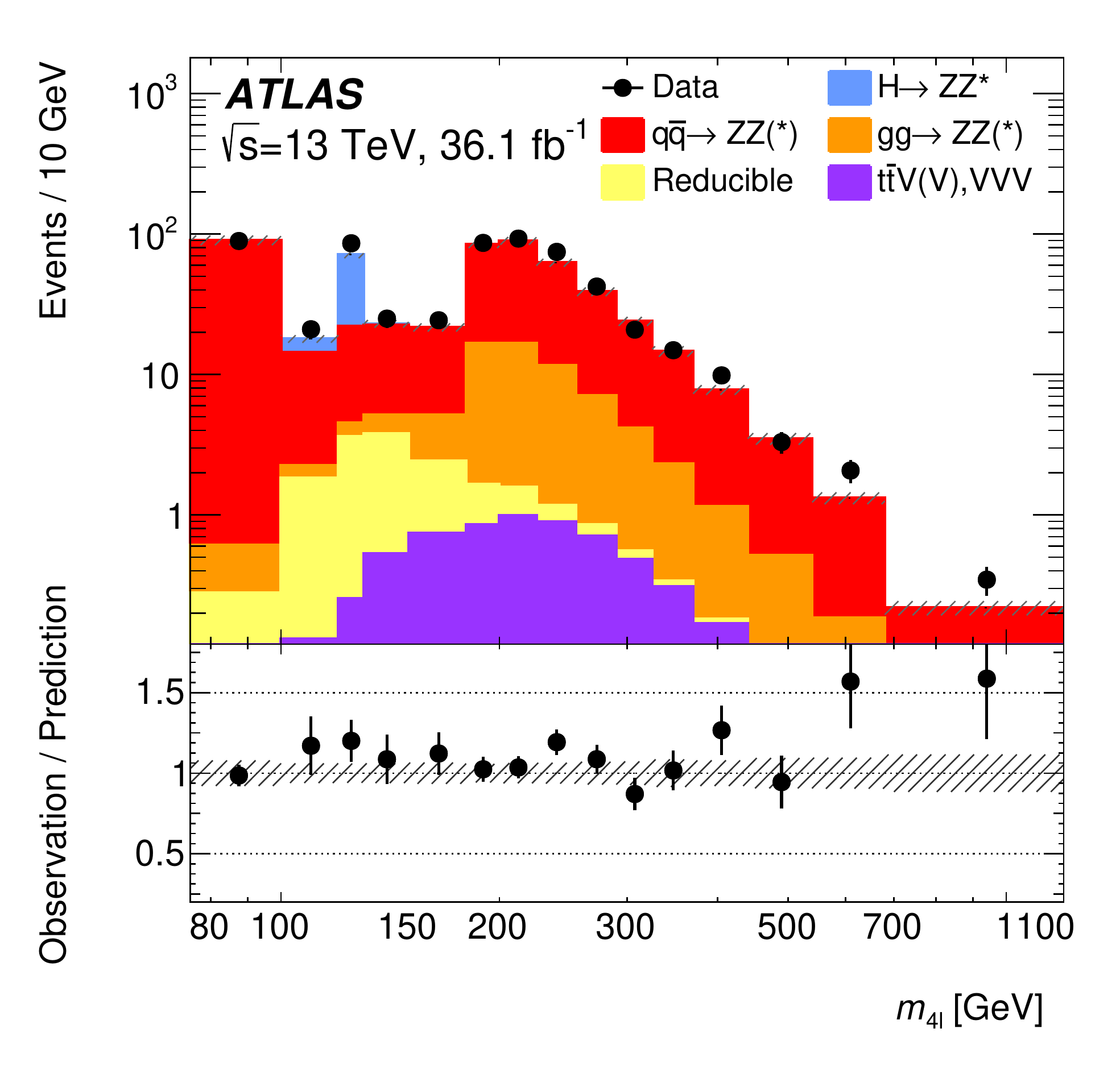}}
        \caption{Distribution of events passing the selection as a function of the four-lepton invariant mass $\mFourL$, where observed event yields (black dots) are compared with the total SM prediction. The ratio of the data to the prediction is given in the lower panel. The statistical uncertainty of the data is displayed with black error bars and the total uncertainty (including statistical and systematic sources) of the prediction is displayed with a grey hashed band.  
        \label{fig:recoresults1}}
    \end{figure}

\begin{figure}[h]
    \centering
       \includegraphics[width = 0.49 \textwidth]{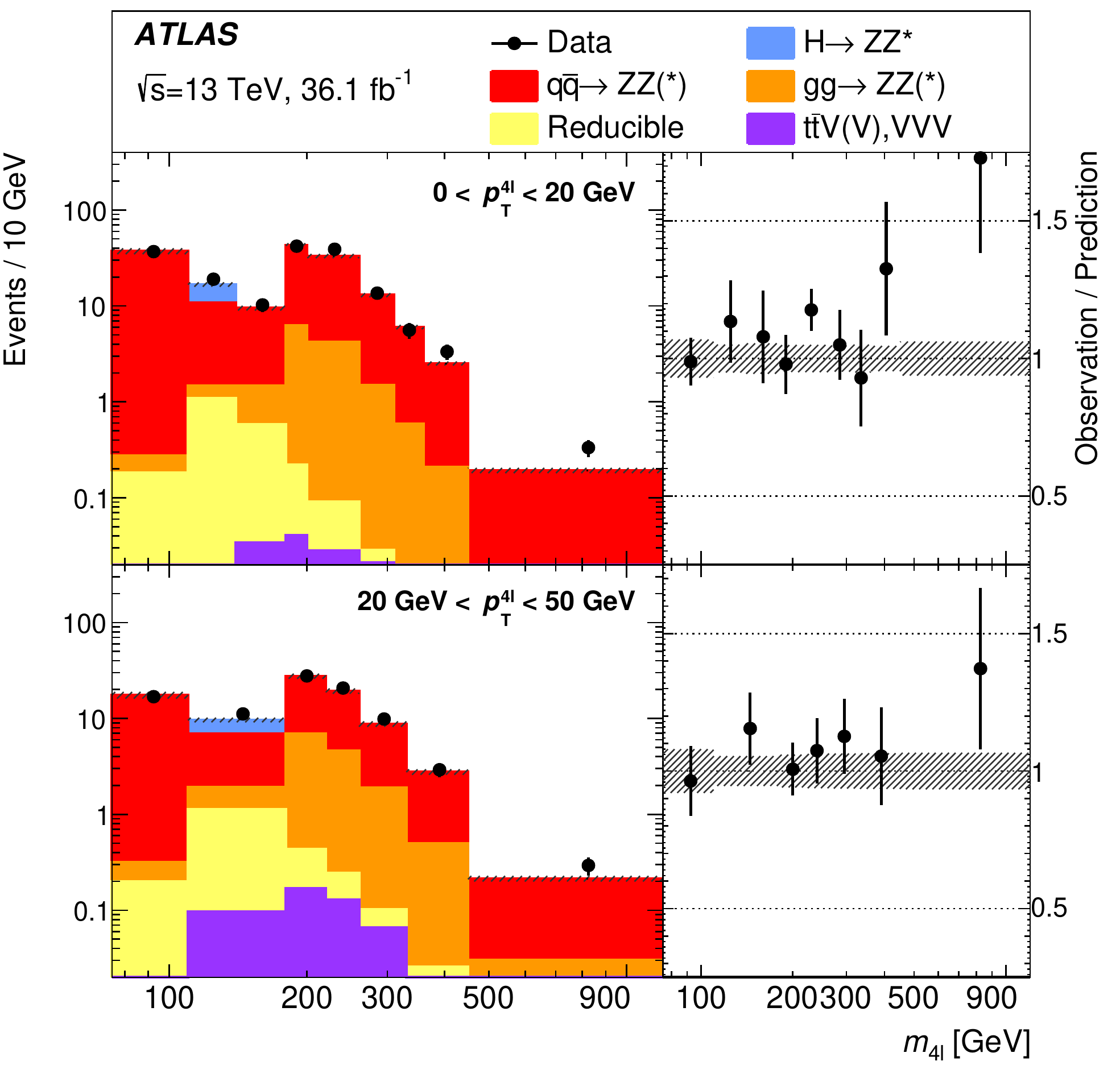}
       \includegraphics[width = 0.49 \textwidth]{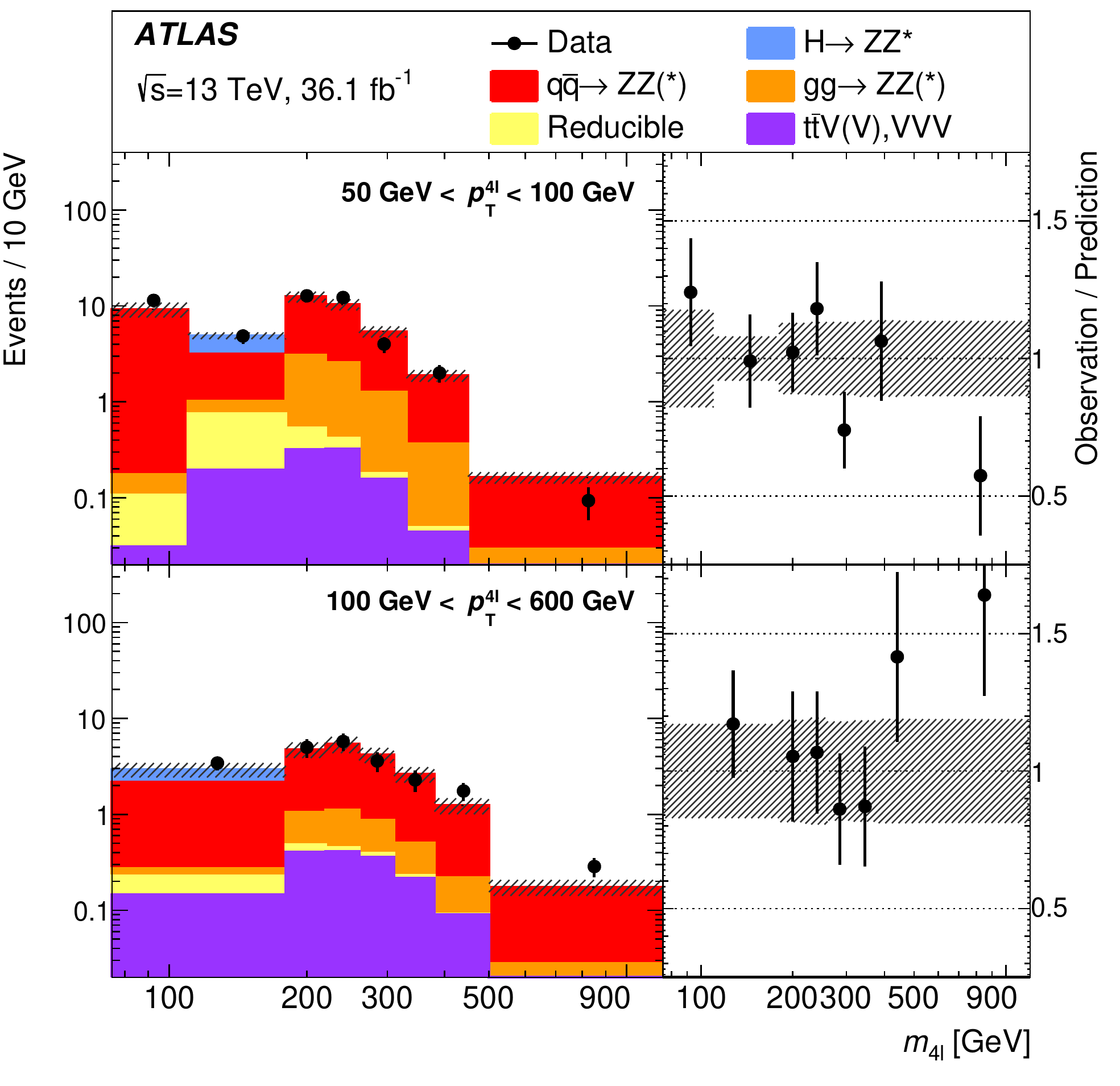}
        \caption{Distribution of events passing the selection as a function of the four-lepton invariant mass $\mFourL$ and of $\ptFourL$, where observed event yields (black dots) are compared with the total SM prediction. The $\mFourL$ bins are shown along the horizontal axis, and the bins of $\ptFourL$ are stacked vertically and labelled with the bin range values. The ratio of the data to the prediction as a function of $\mFourL$ for each secondary variable bin is given in the panel to the right-hand side. The statistical uncertainty of the data is displayed with black error bars and the total uncertainty (including statistical and systematic sources) of the prediction is displayed with a grey hashed band.
         \label{fig:recoresults2a}}
    \end{figure}

\begin{figure}[h]
        \centering
           \includegraphics[width = 0.49 \textwidth]{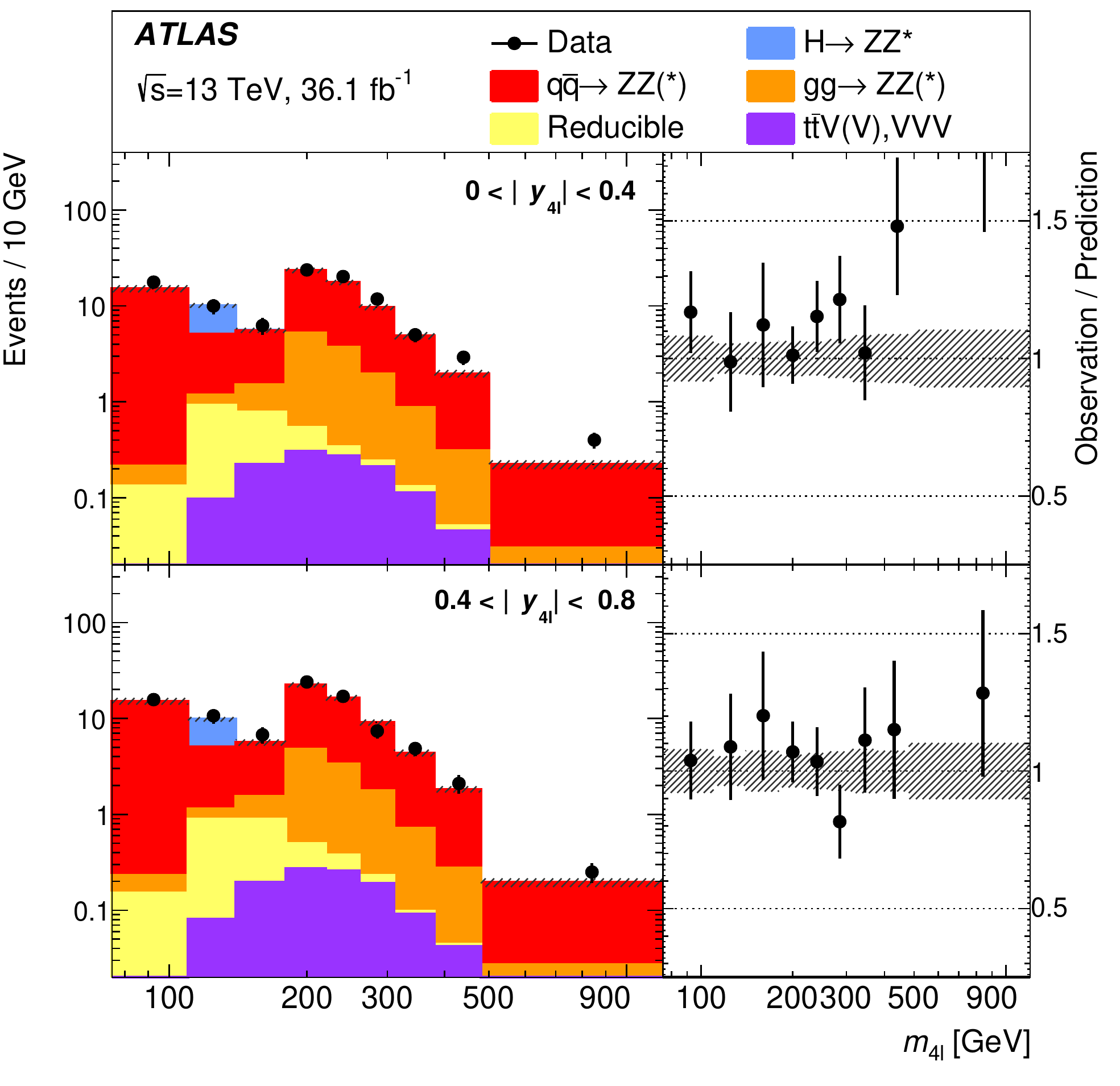}
           \includegraphics[width = 0.49 \textwidth]{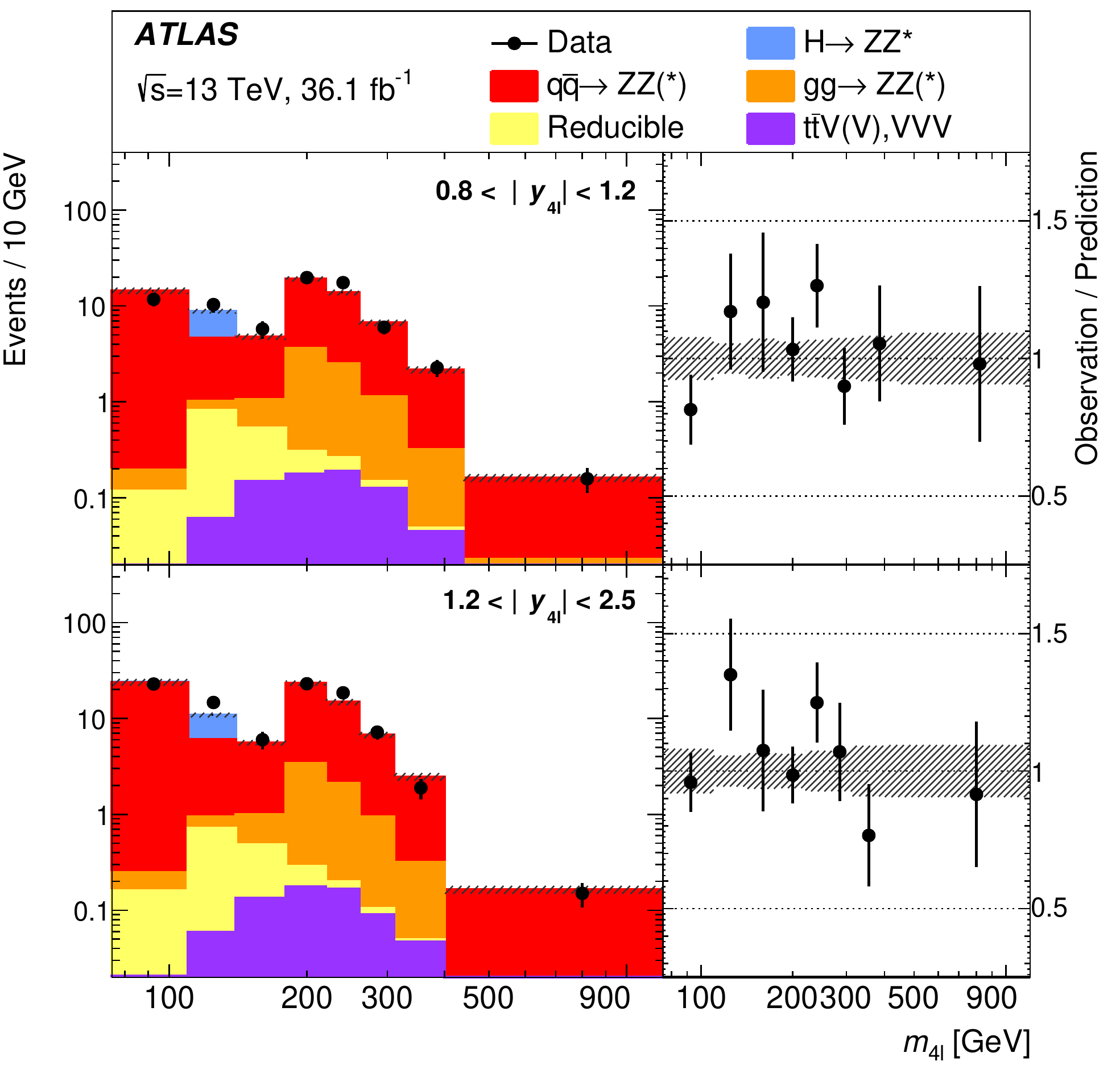}
            \caption{Distribution of events passing the selection as a function of the four-lepton invariant mass $\mFourL$ and of $|\yFourL|$, where observed event yields (black dots) are compared with the total SM prediction. The $\mFourL$ bins are shown along the horizontal axis, and the bins of $|\yFourL|$ are stacked vertically and labelled with the bin range values. The ratio of the data to the prediction as a function of $\mFourL$ for each secondary variable bin is given in the panel to the right-hand side. The statistical uncertainty of the data is displayed with black error bars and the total uncertainty (including statistical and systematic sources) of the prediction is displayed with  a grey hashed band.
             \label{fig:recoresults2b}}
        \end{figure}

\begin{figure}[h]
    \centering
        \subfloat[double-differential $D_\textrm{ME}$--$\mFourL$]{\includegraphics[width = 0.49 \textwidth]{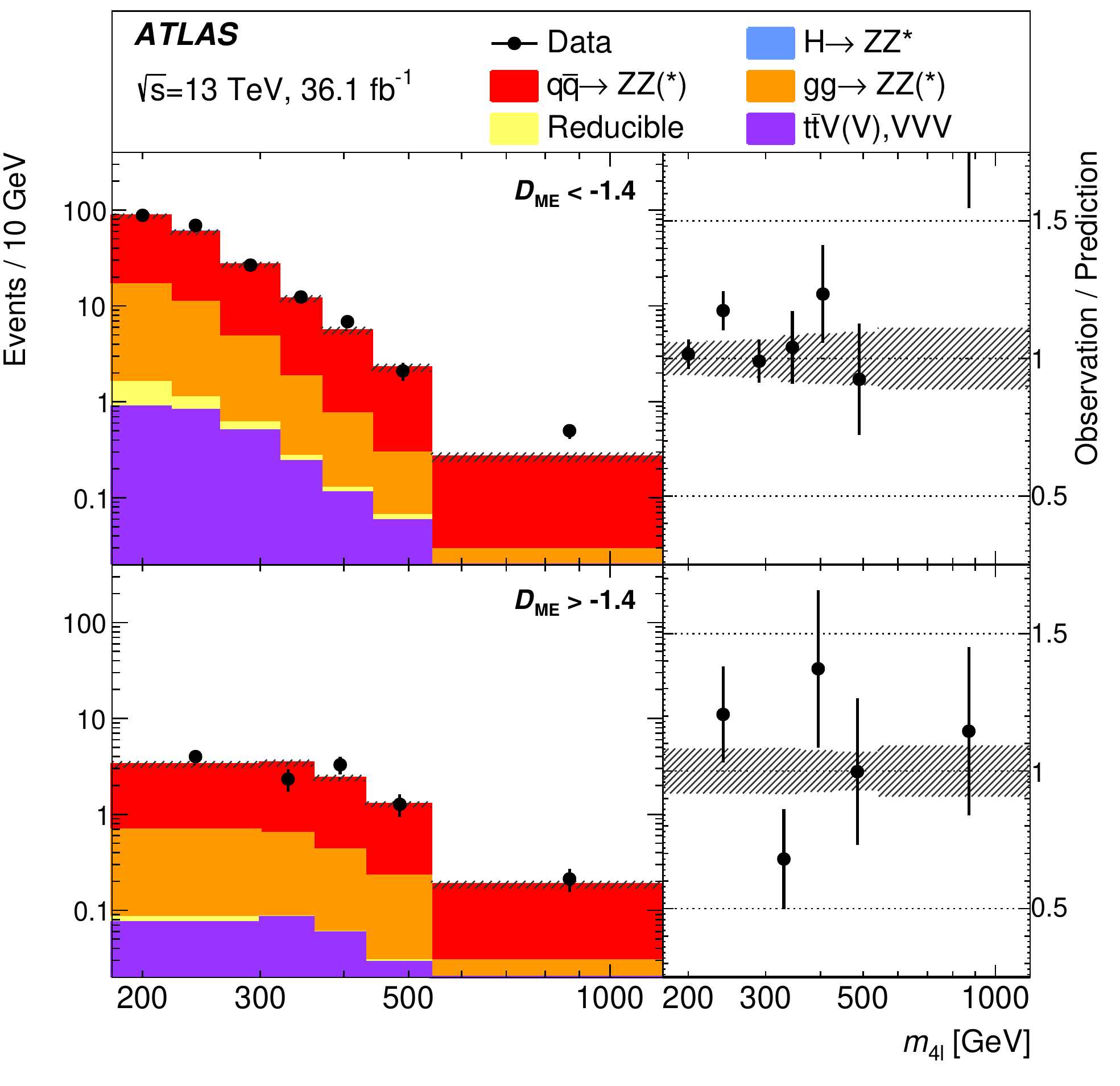}}
        \subfloat[$\mFourL$ per lepton flavour channel]{\includegraphics[width = 0.49 \textwidth]{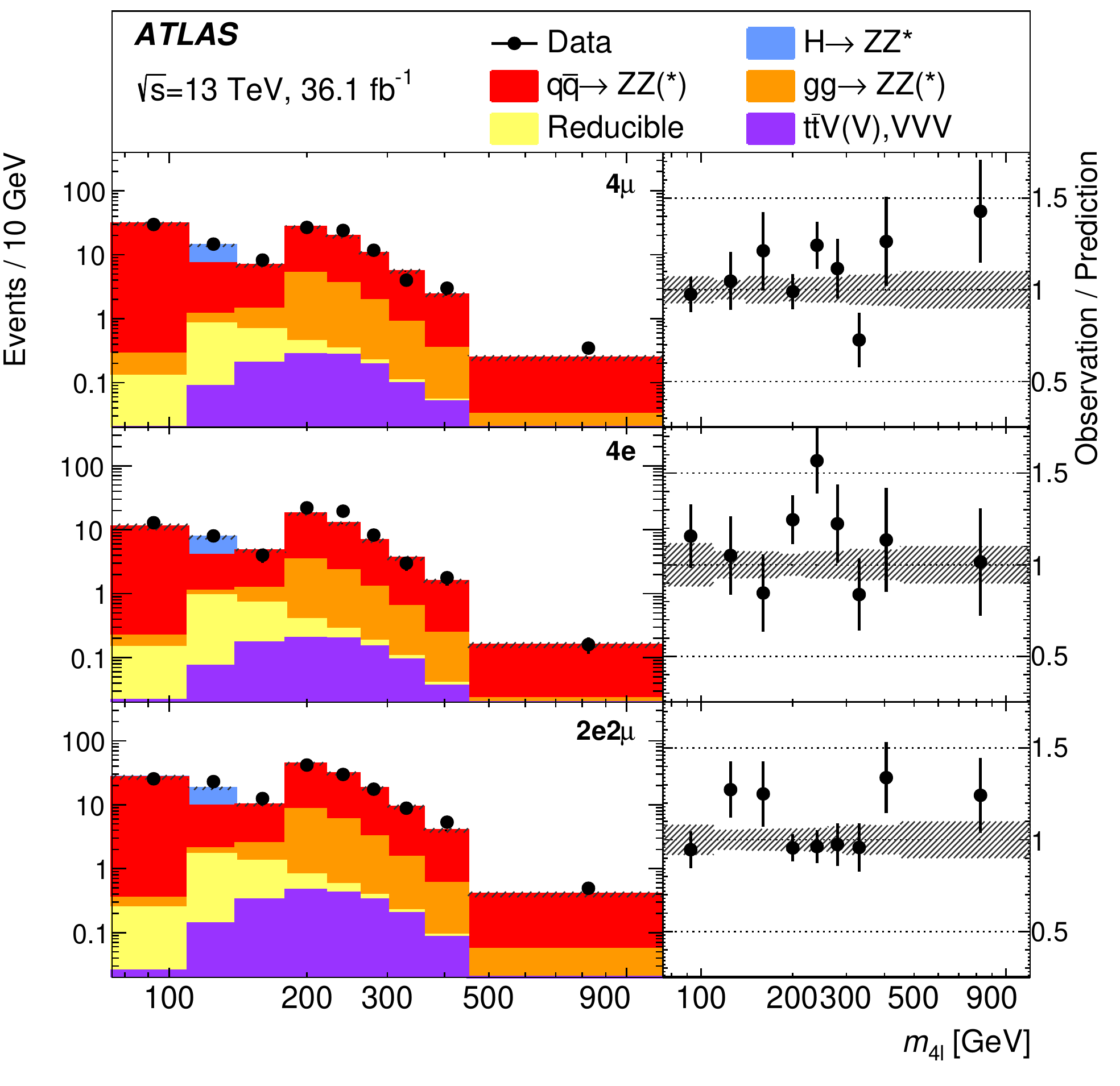}}
        \caption{Distribution of events passing the selection as a function of the four-lepton invariant mass $\mFourL$ and of $D_\textrm{ME}$ (a) and the final-state lepton flavour channel (b), where observed event yields (black dots) are compared with the total SM prediction. The $\mFourL$ bins are given along the horizontal axis, and the bins of the secondary variable  are stacked vertically and labelled with the bin range values. The ratio of the data to the prediction as a function of $\mFourL$ for each secondary variable bin is given in the panel to the right-hand side. The statistical uncertainty of the data is displayed with black error bars and the total uncertainty (including statistical and systematic sources) of the prediction is displayed with  a grey hashed band.
        \label{fig:recoresults3}}
    \end{figure}

Figures~\ref{fig:recoresults1}--\ref{fig:recoresults3} show the observed distributions for events passing the full selection at detector level, 
before unfolding, compared with the expected distributions based on the simulated signal and irreducible background and estimated reducible background processes. 
In the $\mFourL$ distribution, enhancements in the first and third bins correspond to single $Z$ boson production and radiative decay, and on-shell Higgs production, respectively.
An enhancement at around 180~\GeV\ due to the onset of on-shell $ZZ$ production is also clearly visible. 
Overall, no significant discrepancy between the prediction and observation is found. 

\begin{figure}[h]
   \centering
        {\includegraphics[width = 0.72 \textwidth]{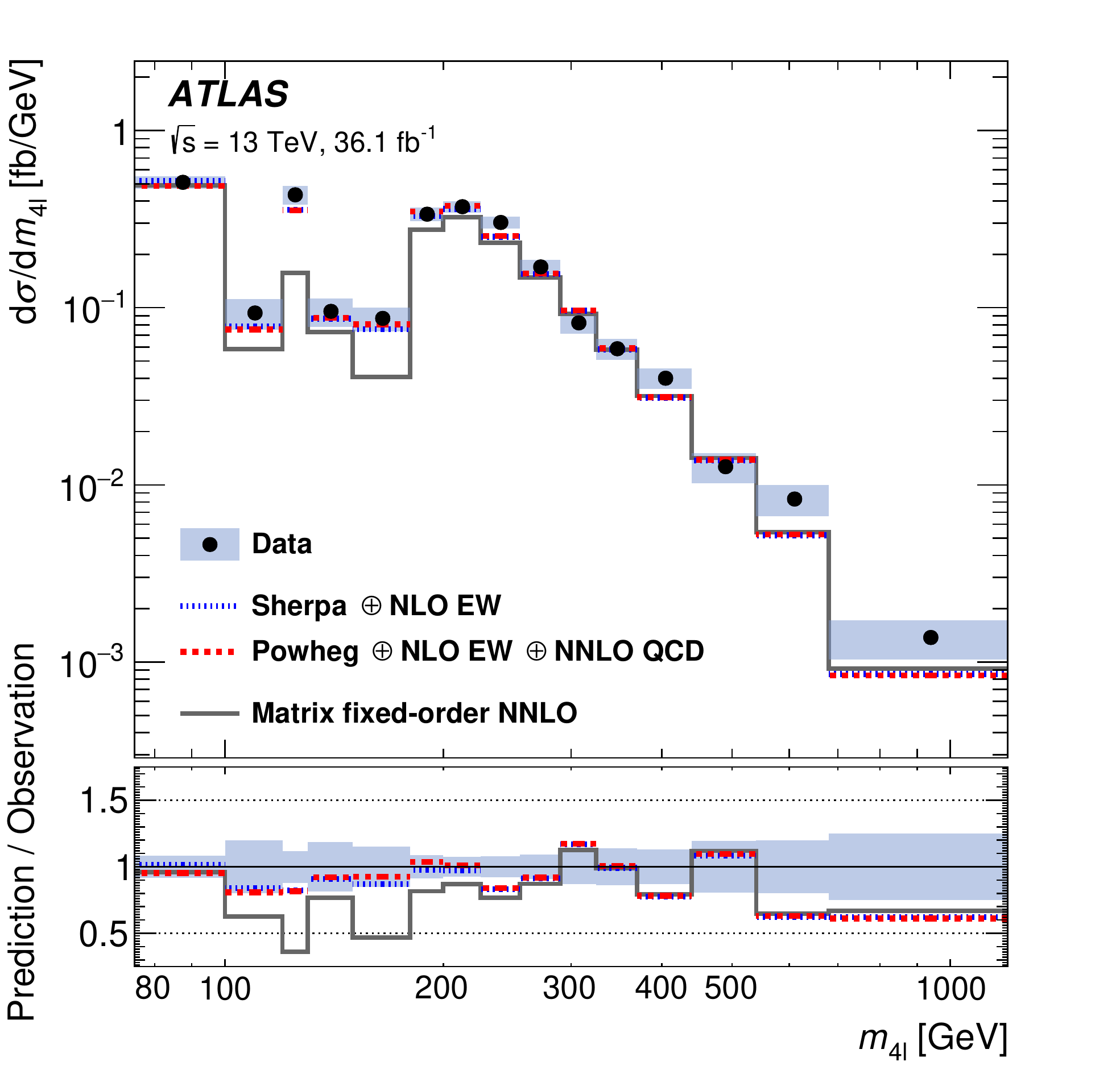}}
        \caption{Measured differential cross-section (black dots) compared with particle-level SM predictions (coloured lines) for the $\mFourL$ distribution. The total systematic plus statistical uncertainty of the measured cross-section is displayed as a grey band. Two SM predictions with different event generator samples for $\qqZZ$ (described in Section~\ref{sec:MC}) are shown with different line colours and styles. In addition, an unmodified NNLO-precision fixed-order calculation using the MATRIX program is shown with a grey histogram, to illustrate the effects of additional higher-order corrections and QED final state radiation included in the event generator predictions. The ratio of the particle-level MC predictions to the unfolded data is shown in the lower panel. 
            \label{fig:unfoldresults}}
    \end{figure}

\begin{figure}[h]
    \centering
        \includegraphics[width = 0.72 \textwidth]{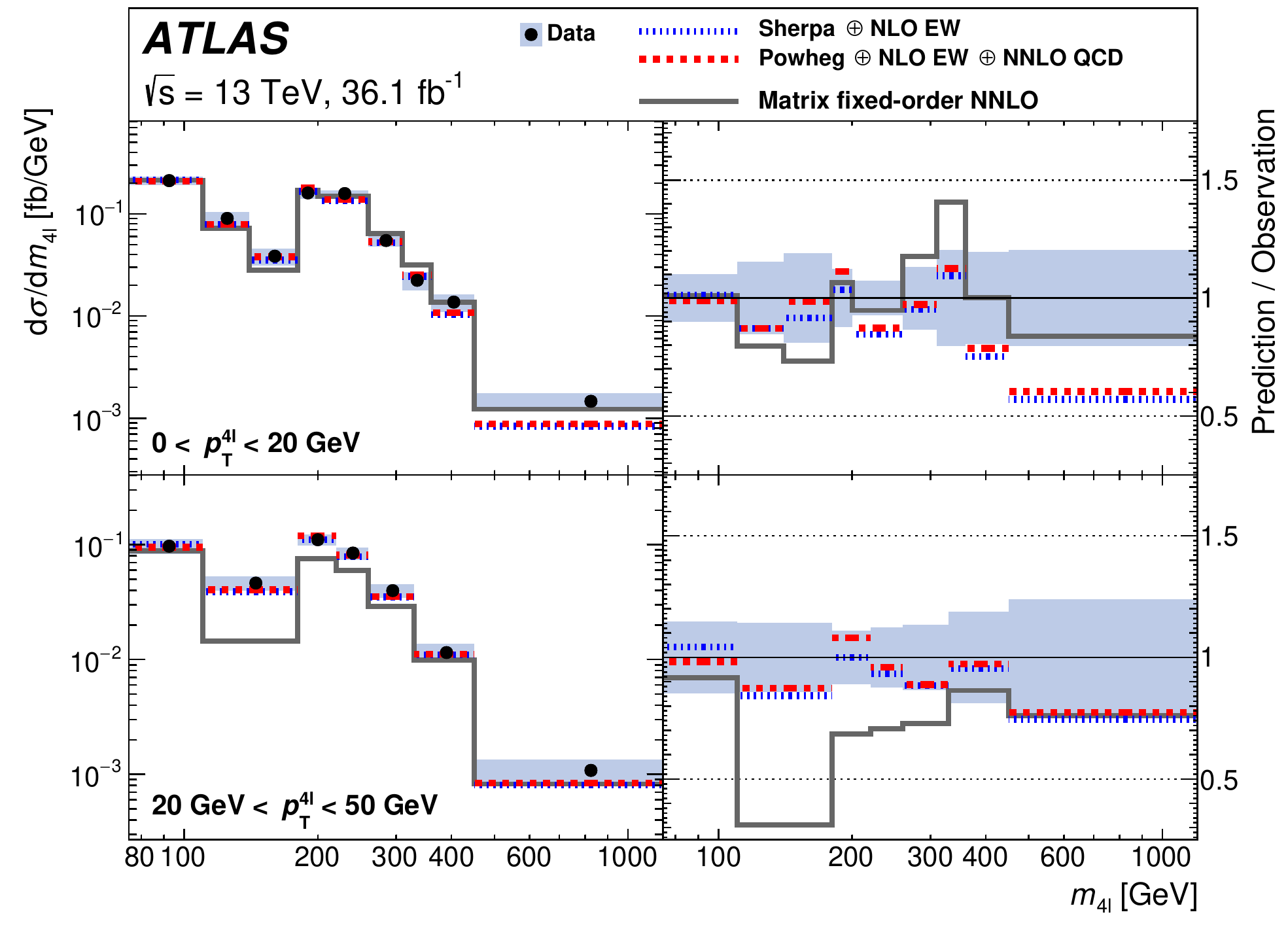}\\
        \includegraphics[width = 0.72 \textwidth]{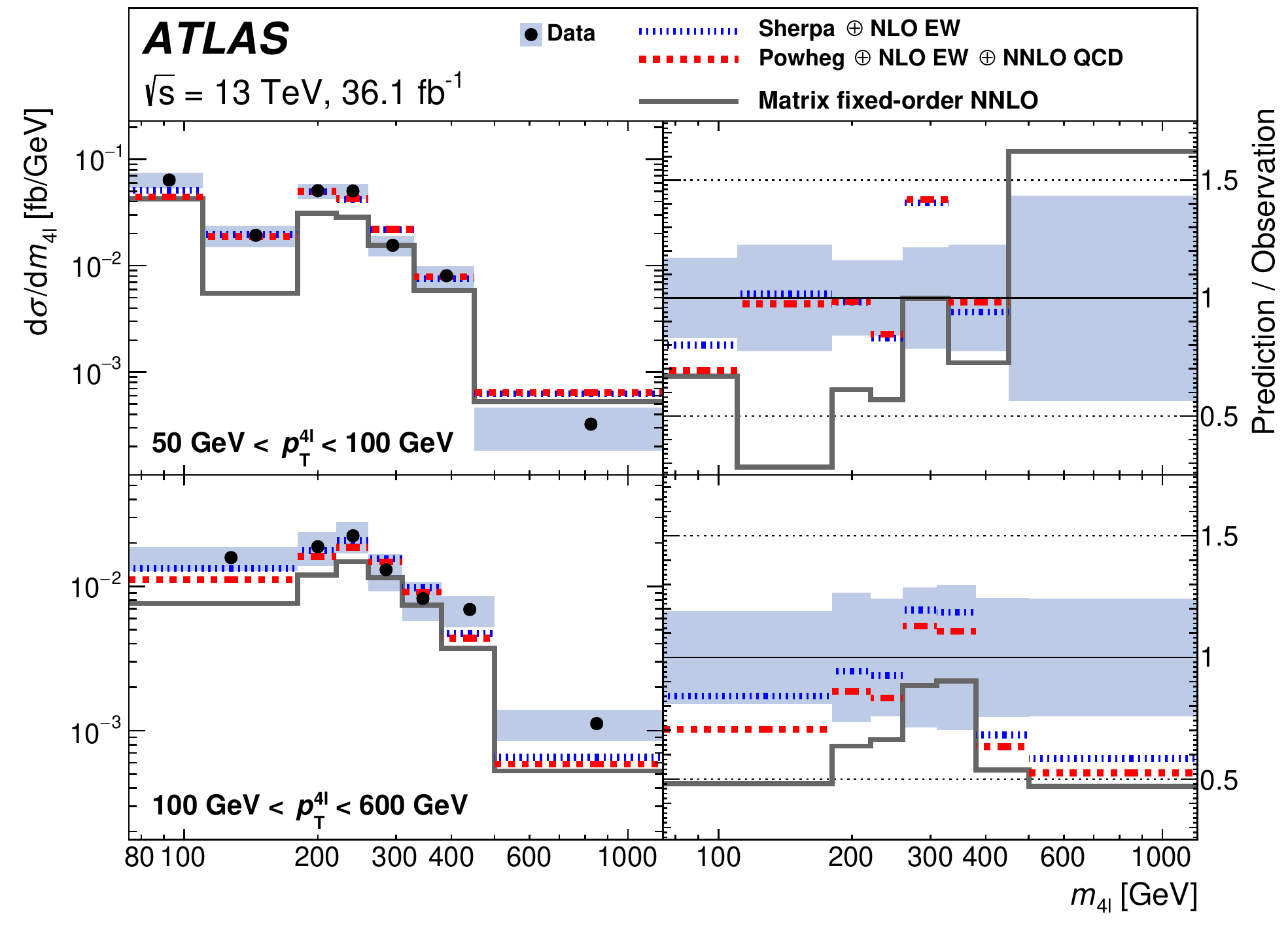}
 \caption{Measured differential cross-section (black dots) compared with particle-level SM predictions (coloured lines) as a function of $\mFourL$ in slices of $\ptFourL$. The total systematic plus statistical uncertainty of the measured cross-section is displayed as a grey band. Two SM predictions with different event generator samples for $\qqZZ$ (described in Section~\ref{sec:MC}) are shown with different line colours and styles. In addition, an unmodified NNLO-precision fixed-order calculation using the MATRIX program is shown with a grey histogram, to illustrate the effects of additional higher-order corrections and QED final state radiation included in the event generator predictions.The $\mFourL$ bins are given along the horizontal axis, and the bins of the secondary variable are stacked vertically and labelled with the bin range values. The ratio of the particle-level MC predictions to the unfolded data as a function of $\mFourL$ for each secondary variable bin is given in the panel to the right-hand side.
            \label{fig:unfoldresultsx}}
    \end{figure}

\begin{figure}[h]
        \centering
            \includegraphics[width = 0.72 \textwidth]{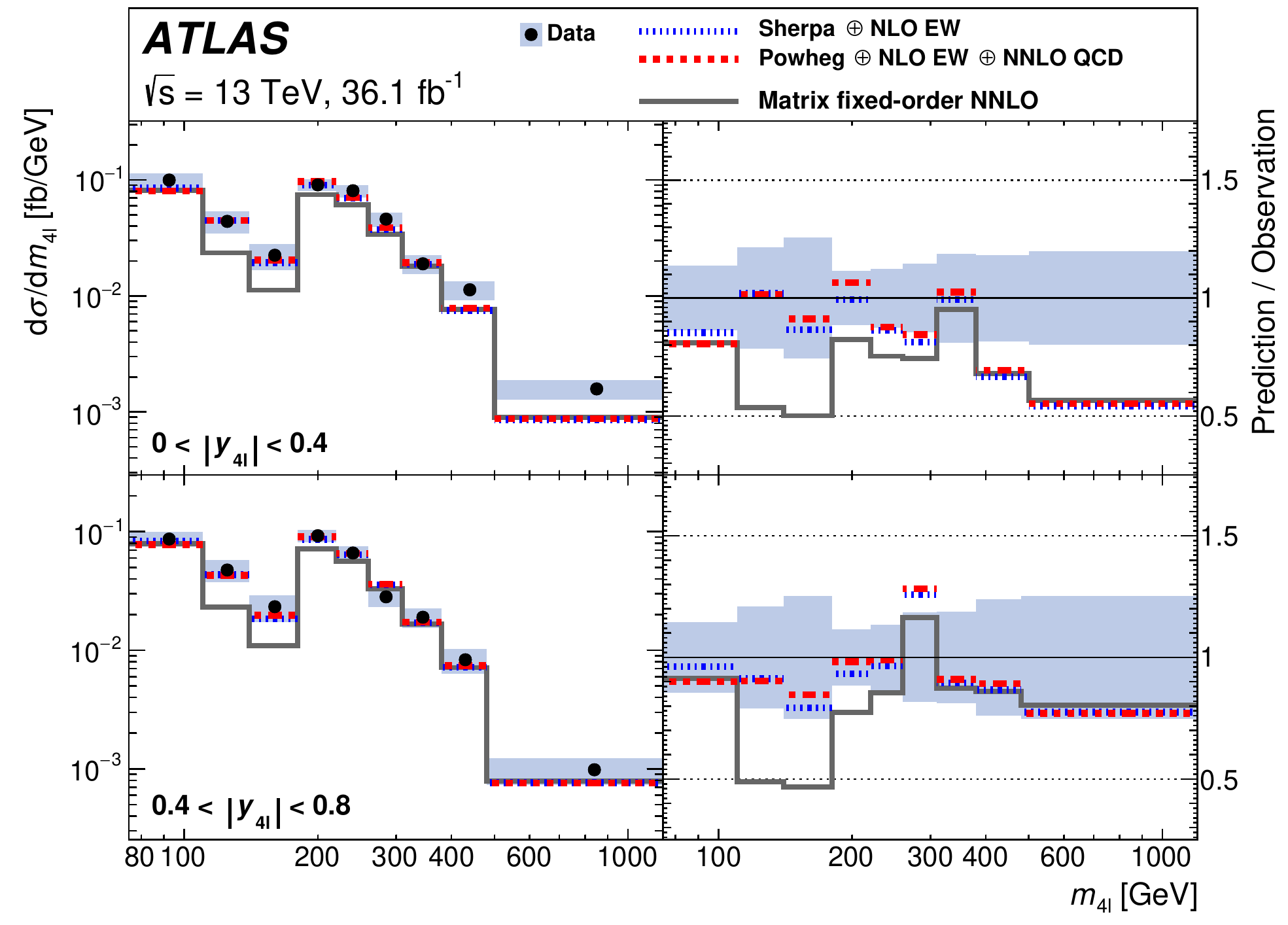}\\
            \includegraphics[width = 0.72 \textwidth]{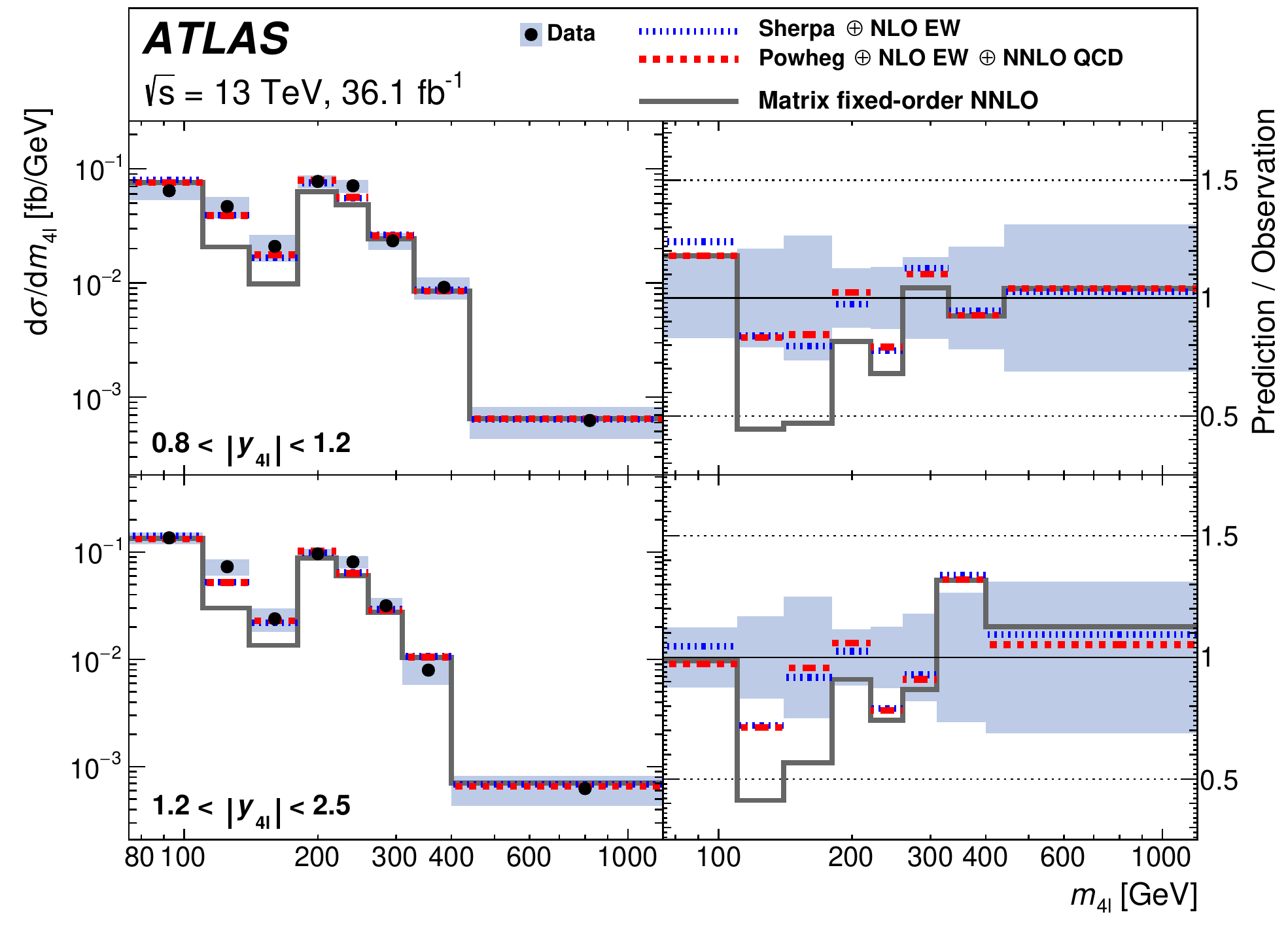}
     \caption{Measured differential cross-section (black dots) compared with particle-level SM predictions (coloured lines) as a function of $\mFourL$ in slices of $|\yFourL|$. The total systematic plus statistical uncertainty of the measured cross-section is displayed as a grey band. Two SM predictions with different event generator samples for $\qqZZ$ (described in Section~\ref{sec:MC}) are shown with different line colours and styles. In addition, an unmodified NNLO-precision fixed-order calculation using the MATRIX program is shown with a grey histogram, to illustrate the effects of additional higher-order corrections and QED final state radiation included in the event generator predictions.The $\mFourL$ bins are given along the horizontal axis, and the bins of the secondary variable are stacked vertically and labelled with the bin range values. The ratio of the particle-level MC predictions to the unfolded data as a function of $\mFourL$ for each secondary variable bin is given in the panel to the right-hand side.
                \label{fig:unfoldresultsy}}
        \end{figure}

\begin{figure}[h]
    \centering
        \includegraphics[width = 0.72 \textwidth]{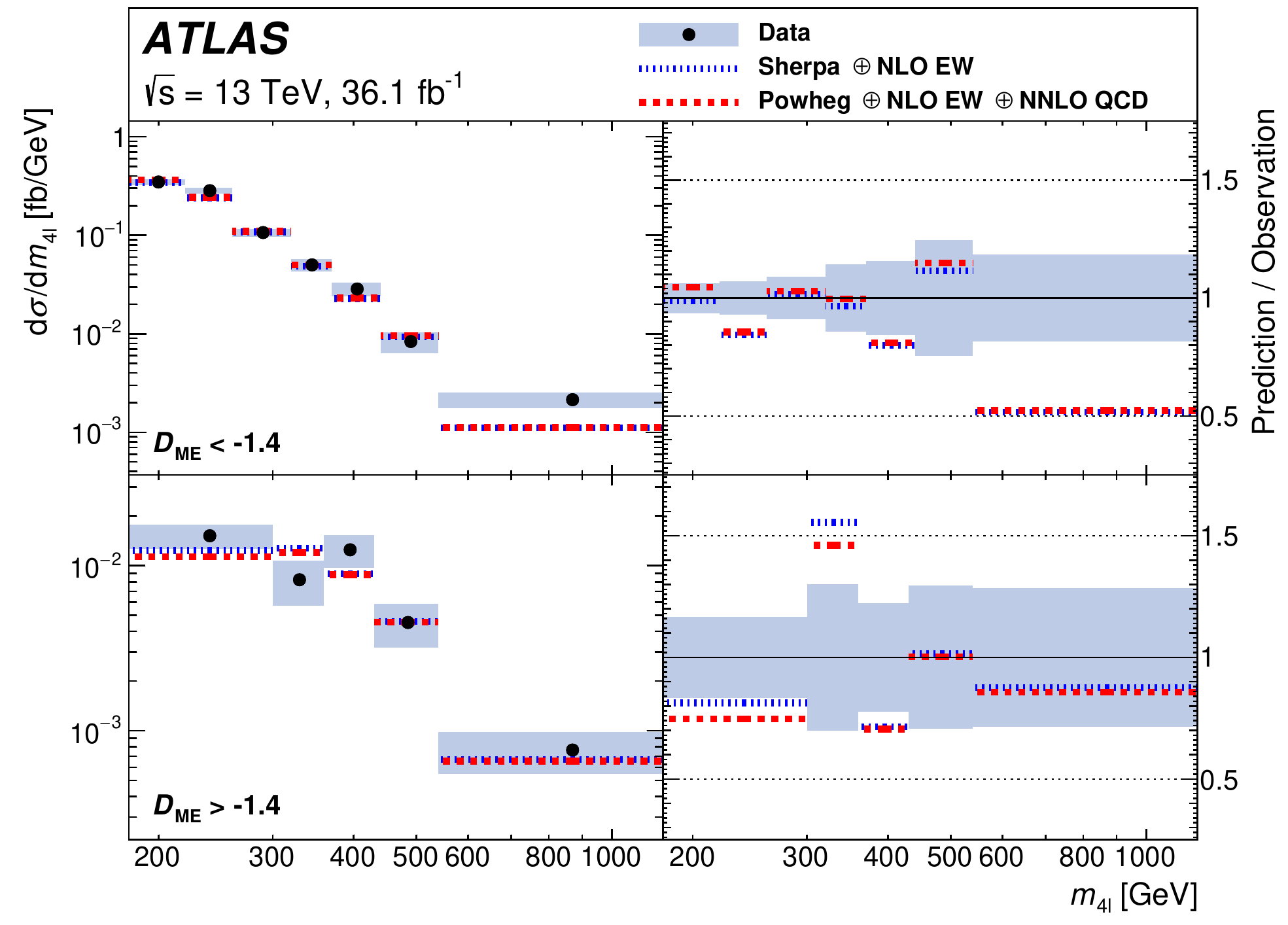}
        \caption{Measured differential cross-section (black dots) compared with particle-level SM predictions (coloured lines) as a function of $\mFourL$ in slices of the $D_\textrm{ME}$ discriminant. The total systematic plus statistical uncertainty of the measured cross-section is displayed as a grey band. Two SM predictions with different generator samples for $\qqZZ$ (described in Section~\ref{sec:MC}) are shown with different line colours and styles. The $\mFourL$ bins are given along the horizontal axis, and the bins of the secondary variable are stacked vertically and labelled with the bin range values. The ratio of the particle-level MC predictions to the unfolded data as a function of $\mFourL$ for each secondary variable bin is given in the panel to the right-hand side.
            \label{fig:unfoldresultsmela}}
    \end{figure}

\begin{figure}[h]
    \centering
        \includegraphics[width = 0.72 \textwidth]{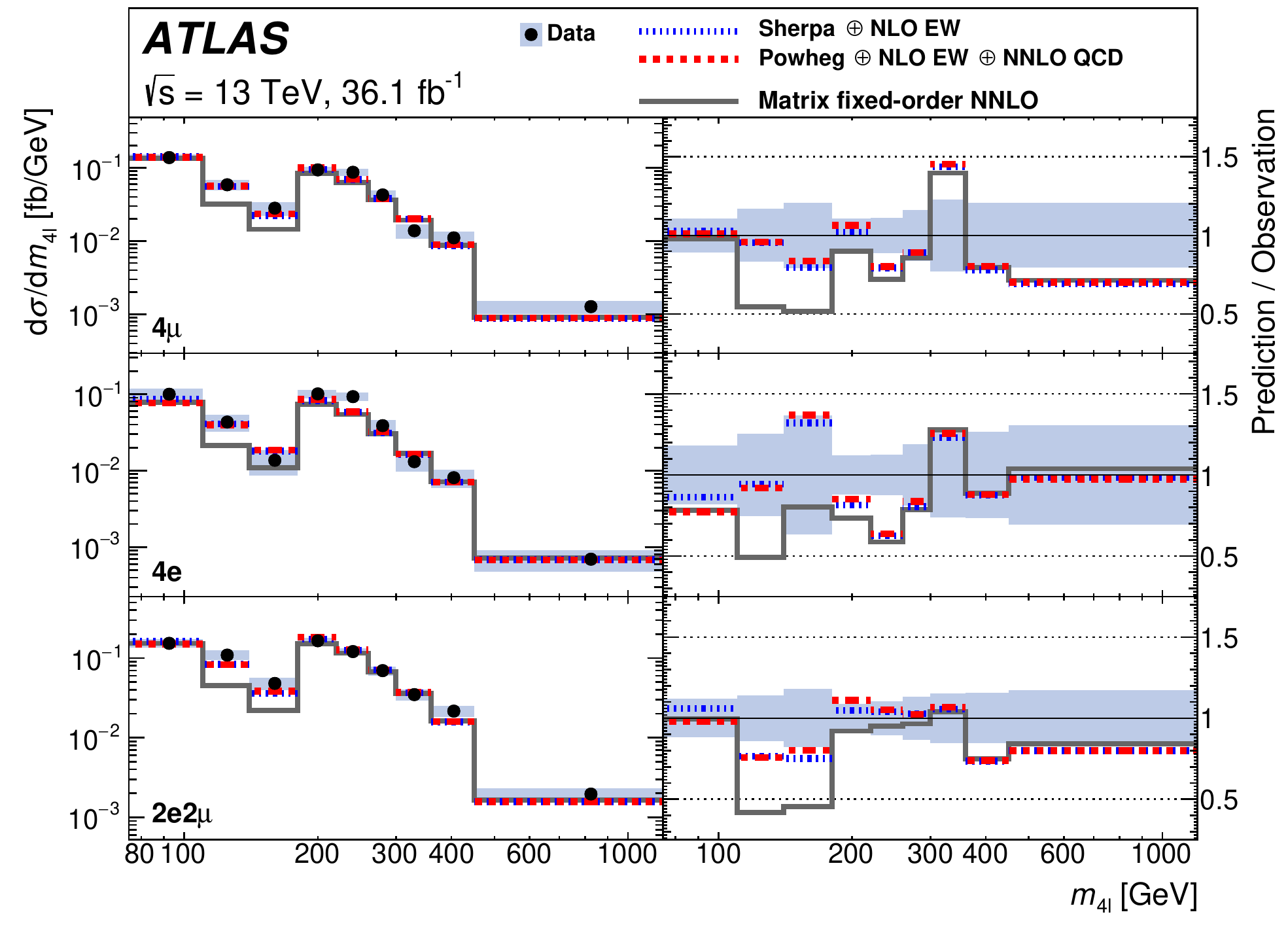}\\
        \caption{Measured differential cross-section (black dots) compared with particle-level SM predictions (coloured lines) as a function of $\mFourL$ for each final-state lepton flavour configuration. The total systematic plus statistical uncertainty of the measured cross-section is displayed as a grey band. Two SM predictions with different generator samples for $\qqZZ$ (described in Section~\ref{sec:MC}) are shown with different line colours and styles. In addition, an unmodified NNLO-precision fixed-order calculation using the MATRIX program is shown with a grey histogram, to illustrate the effects of additional higher-order corrections and QED final state radiation included in the event generator predictions. The $\mFourL$ bins are given along the horizontal axis, and the bins of the secondary variable are stacked vertically and labelled with the bin range values. The ratio of the particle-level MC predictions to the unfolded data as a function of $\mFourL$ for each secondary variable bin is given in the panel to the right-hand side.
            \label{fig:unfoldresultsch}}
    \end{figure}

The observed distributions are then corrected for detector effects by unfolding as described in Section~\ref{sec:unfolding}. 
The resulting measured differential cross-section as a function of $\mFourL$ and double-differential cross-sections as functions of $\mFourL$ and $\ptFourL$, $|\yFourL|$, the $D_\textrm{ME}$ discriminant, or the final-state lepton flavour configuration are shown in Figures~\ref{fig:unfoldresults}--\ref{fig:unfoldresultsch}, and compared with particle-level predictions.

Overall the predictions are consistent with the measurement when using either $\SHERPAV{2.2.2}$ or $\POWHEG+\PYTHIAV{8}$ to describe the dominant $\qqZZ$ component, considering the systematic and statistical uncertainties. 

Furthermore, the predictions from $\SHERPAV{2.2.2}$ and $\POWHEG+\PYTHIAV{8}$ are in excellent agreement. This gives confidence in the validity of the procedure used to reweight $\POWHEGBOX$ events to NNLO QCD accuracy by applying $\mFourL$-based $K$-factors calculated with MATRIX~\cite{Cascioli:2014yka,Grazzini:2015hta,Grazzini:2017mhc,Kallweit:2018nyv}. It also indicates that, at least for this observable, an analogous reweighting of $\SHERPA$ events is not required due to this generator's intrinsic higher accuracy.
The fixed-order NNLO QCD prediction by MATRIX shows an expected underestimation at and below the on-shell $m_{ZZ}$ threshold. This underestimation is mainly due to missing real, wide-angle QED emission effects in events where both $Z$ bosons are on-shell, and amounts to several tens of percent of the total population in the region just below the on-shell threshold~\cite{Biedermann:2016lvg}. For the $\SHERPAV{2.2.2}$ and $\POWHEG+\PYTHIAV{8}$ samples, QED effects are included from estimates taken from QED shower programs. Moreover, the fixed-order MATRIX prediction is equivalent to having leading-order precision for the continuum \ggZZ process and on-shell Higgs boson production, while the event generator samples include sizeable higher-order contributions.
The predictions from $\SHERPA$, $\POWHEGBOX$ and MATRIX agree at the level of a few percent, outside the region of resonant Higgs boson production, if the comparison is performed prior to QED showering and without both the additional NLO electroweak corrections and the application of higher-order corrections to the $\ggZZ$ contribution. 

    \FloatBarrier

\section{Interpretations}
\label{sec:interpretations}
The measured particle-level differential and double-differential fiducial cross-sections can be interpreted to measure SM parameters and set limits on BSM contributions. 
To explore and demonstrate this potential, a range of interpretations are presented in this paper. 
The production rate of \ggB{} is extracted with respect to the SM prediction using the differential cross-section measured as a function of $\mFourL$. 
The $Z\rightarrow 4 \ell$ branching fraction is estimated from the measured fiducial cross-section in the mass bin corresponding to $m_{Z}$. 
Constraints on the rate of off-shell Higgs boson production ($gg \rightarrow H^{\ast} \rightarrow 4\ell$) are derived using the double-differential cross-section measured as a function of $\mFourL$ and the $D_\textrm{ME}$ discriminant, which greatly enhances sensitivity to this type of process. 
Constraints on modified couplings of the Higgs boson to top quarks and gluons in the off-shell region are also derived, using the measured differential cross-section as a function of $\mFourL$. 

All interpretations use a common statistical approach. 
A multivariate Gaussian likelihood function is used to quantify the level of agreement between a given prediction and observed data simultaneously across all bins of a measurement, taking into account correlations due to bin migration.
The $\chi^2$ function defining the exponential component of the likelihood takes the form:
\begin{equation*}
\chi^{2} = (y_{\text{data}} - y_{\text{pred}})^\text{T}C^{-1}(y_{\text{data}} - y_{\text{pred}}),
\end{equation*} 
where $y_{\text{data}}$ is a vector of unfolded observed values in each of the distribution bins, $y_{\text{pred}}$ is a vector of the predicted values in 
each of the distribution bins, which is a function of the parameter of interest (POI) and nuisance parameters (NP), and $C^{-1}$ is the inverse of the total covariance matrix for the prediction being tested. 
This covariance matrix is obtained by rescaling the covariance matrix resulting from unfolding the detector-level SM prediction, to account for the change in the predicted yield relative to the original prediction for the values of the POI and NP under consideration. 
Each element $C(i,j)$ of the rescaled matrix corresponding to bins $i$ and $j$ can be expressed using the systematic, statistical and background components $C_{\text{syst}}^\text{SM}$, $C_{\text{stat}}^\text{SM}$ and $C_{\text{bkg}}^\text{SM}$ of the covariance matrix corresponding to the SM prediction:
\begin{equation*}
C(i,j) = R_{i}\times R_{j}\times C_{\text{syst}}^\text{SM}(i,j) + \sqrt{(R_{i}\times R_{j})} \times C_{\text{stat}}^\text{SM}(i,j) + C_{\text{bkg}}^\text{SM}(i,j),
\end{equation*}
where $R_{k}= N^{\text{pred}}_{k}(\text{{POI, NP}}) / N^{\text{pred}}_{k}(\text{POI}=\text{SM}, \text{NP}=0)$ is the ratio of the predicted yield in bin $k$ assuming the given values of the parameter of interest and nuisance parameters to the yield in bin $k$ using the SM value of the POI and a nominal value of the NP. 
All sources of experimental uncertainty, including those related to the unfolding procedure itself, are included in the systematic covariance matrix. 
The background component includes any uncertainties in the estimated background subtracted prior to unfolding and does not vary with the POI or NP. 
Theoretical uncertainties in the predictions do not enter the covariance matrix but are modelled with a nuisance parameter for each of the shape 
and normalisation components, constrained with Gaussian probability density functions. 

Upper limits on the values of the parameters of interest are set using the CL$_\text{s}$ method~\cite{CLsMethod} with a confidence level of 95$\%$.    

\subsection*{Signal strength for gluon-induced $4\ell$ production}

The best prediction for the fiducial cross-section for gluon-induced  $4\ell$ production (\ggB) in the interval $180~\GeV<\mFourL<1200~\GeV$, where the Higgs resonance is not dominant, is approximately 6.5~fb, compared to a leading order \MCFM prediction of 3.0~fb. 
The relative contribution of $\ggZZ$ to the differential $pp\rightarrow 4\ell$ cross-section is greatest in the region $180~\GeV{} < \mFourL \lesssim 400~\GeV$, contributing around $18\%$ at $\mFourL \sim 200~\GeV$, as visible in Figure~\ref{fig:recoresults1}. 
For a comparison with the best theoretical prediction, the signal strength for this process, $\mu_{gg} = \sigma^{\text{measured}}_{\ggB} /\sigma^{\text{SM}}_{\ggB}$, is extracted. 
The differential $\mFourL$ distribution is used for this interpretation, as NLO QCD precision is available in the description of this variable. 
A likelihood scan is performed using the procedure outlined above. The contribution from $\qqZZ$ production is set to the theoretical prediction as described in Section~\ref{sec:MC} and allowed to vary within the associated theoretical uncertainties described in Section~\ref{sec:uncertainties} by means of nuisance parameters with Gaussian constraints. 
The best available simulation of $gg \rightarrow 4\ell$ as described in Section~\ref{sec:MC} is scaled by the parameter of interest, $\mu_{gg}$, and in addition also allowed to vary within the associated theoretical uncertainties. 
A signal strength $\mu_{gg} = 1.3 \pm 0.5$ is measured with an expected value of $1.0 \pm 0.4$.
In addition, a signal strength $\mu^{\text{LO}}_{gg} = \sigma^{\text{measured}}_{\ggB} /\sigma^{\text{SM, LO QCD}}_{\ggB}$, is extracted relative to an uncorrected leading-order precision \MCFM prediction of $\ggB$ as $\mu^{\text{LO}}_{gg} = 2.7\pm0.9$, with an expected value of $2.2\pm0.9$. 
This value can be compared with a previous ATLAS measurement of $\mu^{\text{LO}}_{gg}=2.4\pm1.4$ performed at $\sqrt{s}=8$~TeV~\cite{STDM-2014-15}.
In both cases, the uncertainty is dominated by data statistics. The largest systematic uncertainty contribution is the QCD scale choice in the $\qqZZ$ prediction, and is small compared to the statistical uncertainty.  
Consistent results were also obtained when using the double-differential $\mFourL$--$\ptFourL$ or $\mFourL$--$\yFourL$ distributions and the $\mFourL$ measurement per final-state flavour configuration, all of which showed comparable sensitivity.

\subsection*{Extraction of the $Z\rightarrow 4\ell$ branching fraction}
The branching fraction of $Z\rightarrow 4 \ell$ is extracted using the lowest $\mFourL$ bin (75--100~\GeV) in the unfolded $\mFourL$ distribution shown in Figure~\ref{fig:unfoldresults}.
This bin is dominated by single $Z$ boson production (Figure~\ref{fig:ZZproduction}(c)), but there are minor non-resonant contributions from $t$-channel $q\bar{q}$ production (Figure~\ref{fig:ZZproduction}(a)) and $gg\rightarrow ZZ^{(*)}$ (Figure~\ref{fig:ZZproduction}(b)).
The measurement is performed in an extended phase space defined by values of the invariant mass of the four-lepton  system $\mFourL$ and the lowest dilepton invariant mass in the event, $m_{\ell\ell}$, satisfying $80 < \mFourL < 100$~\GeV\ and $m_{\ell\ell}>$4~\GeV. 
The branching fraction is then calculated as:
\begin{equation*}
\mathcal{B}_{Z\rightarrow 4\ell} = \frac{N_{\text{fid}} \times (1-f_{\text{non-res}})}{\sigma_{Z} \times A_{\text{fid}} \times \mathcal{L}},
\end{equation*}
where $N_{\text{fid}}$ is the number of unfolded events in this bin, $A_{\text{fid}}$ is the fiducial acceptance, defined as the ratio of the events passing the fiducial selection to those in the extended phase space, $\sigma_{Z}$ is the total cross-section for single $Z$ production, $\mathcal{L}$ is the integrated luminosity, and $f_{\text{non-res}}$ is the fraction of non-resonant events in the extended phase space, calculated using \POWHEGBOX. 
The acceptance (including the non-resonant contribution) is calculated using MC simulation as $A_{\text{fid}} = (4.75\pm0.02)\%$ and the fraction of non-resonant events as $f_{\text{non-res}} =(4.8 \pm 0.5)\% $, where the uncertainty includes the statistical uncertainty of the samples used and the systematic uncertainty from the theoretical variations described in Section~\ref{sec:uncertainties}.

The branching fraction is measured to be
\begin{equation*}
\mathcal{B}_{Z\rightarrow 4\ell} = [4.70\pm0.32(\text{stat})\pm0.21(\text{syst})\pm0.14(\text{lumi})]\times 10^{-6}
\end{equation*}
using the measured value for $\sigma_{Z}$ from Ref.~\cite{STDM-2015-03}.
Here, the systematic uncertainty includes the systematic uncertainty of the measured $\sigma_{Z}$ and the systematic uncertainty of the unfolded cross-section in the bin used for the measurement, as well as the uncertainty in $A_{\text{fid}}$ and $f_{\text{non-res}}$. As Ref.~\cite{STDM-2015-03} is based on $81~\text{pb}^{-1}$ of $pp$ collision data taken during the 2015 LHC run while this measurement uses the full 2015--2016 ATLAS dataset comprising $36~\text{fb}^{-1}$, all detector-related systematic uncertainties as well as the luminosity uncertainty of $\sigma_{Z}$ are conservatively treated as uncorrelated with the equivalent uncertainties in the measured cross-section in the lowest $\mFourL$ bin.

This result is compared with previous dedicated measurements by the ATLAS~\cite{STDM-2013-08} and CMS~\cite{CMS-SMP-16-017} collaborations in Table~\ref{tab:Z4lResultOverview}. The largest contributing systematic uncertainties in this mass region come from lepton identification and reconstruction efficiencies, as shown in Figure~\ref{fig:sysm4l}. The difference in systematic uncertainties compared to Ref.~\cite{STDM-2013-08} is due to the assumptions of non-correlation between uncertainties in the two contributing measurements discussed above. The larger statistical uncertainty compared to Ref.~\cite{CMS-SMP-16-017} arises from an acceptance which has not been fully optimised for this interpretation. Nevertheless, the final precision including all error sources allows this measurement to contribute an improvement in the total precision of the $Z\rightarrow 4\ell$ branching fraction.

\begin{table}[h]

\centering
\caption{ Comparison of measurements for the $Z\rightarrow 4\ell$ branching fraction in the phase-space region $80~\GeV{}~<~\mFourL~<~100~\GeV{}$, $m_{\ell\ell}~>~4~\GeV{}$. \label{tab:Z4lResultOverview}}

\begin{tabular}{l l}
\hline 
\hline 
Measurement  &  \multicolumn{1}{c}{$\mathcal{B}_{Z\rightarrow 4\ell} / 10^{-6}$} \\
\hline 
ATLAS, $\sqrt{s} = 7$~\TeV\ and 8~\TeV~\cite{STDM-2013-08} &  $4.31\pm$0.34(stat)$\pm$0.17(syst)  \\
CMS, $\sqrt{s} = 13$~\TeV~\cite{CMS-SMP-16-017} & $4.83\phantom{.}^{+0.23}_{-0.22}$(stat)$\phantom{.}^{+0.32}_{-0.29}$(syst)$\pm 0.08$(theo)$\pm 0.12$(lumi)   \\

\textbf{ATLAS, $\sqrt{s}=13~\TeV$}   & \textbf{$4.70\pm0.32(\text{stat})\pm0.21(\text{syst})\pm 0.14(\text{lumi})$}   \\
\hline 
\hline 
\end{tabular}
\end{table}

\subsection*{Constraint on off-shell Higgs boson signal strength}
The double-differential distribution for $\mFourL$--$D_\textrm{ME}$ is used to constrain the off-shell Higgs production process at high mass ($\mFourL>$180~\GeV), assuming that the contribution of the box diagram is as predicted by the Standard Model. 
As in the extraction of the signal strength for gluon-induced $4\ell$ production, a likelihood scan is performed where the contribution of $\qqZZ$ is set to the Standard Model prediction and allowed to float within the associated theoretical uncertainties.
The total yield from $\ggZZ$ is then parameterised~\cite{HIGG-2017-06} as 
\begin{equation*}
        N^{\ggZZ}\left(\mu_H^{\text{OS}} \right) = \left(\mu_H^{\text{OS}}-\sqrt{\mu_H^{\text{OS}}}\right) \times N_{\text{SM}}^{\ggS} + \left(1-\sqrt{\mu_H^{\text{OS}}}\right) \times N_{\text{SM}}^{gg\rightarrow 4\ell \text{(box)}}  + \sqrt{\mu_H^{\text{OS}}} \times N_{\text{SM}}^{\ggZZ},
\end{equation*}
where $\mu_H^{\text{OS}} = \sigma_{gg\rightarrow H^\ast \rightarrow 4\ell} / \sigma^{\text{SM}}_{gg\rightarrow H^\ast \rightarrow 4\ell} $ is the signal strength for the off-shell Higgs production process, the parameter of interest for this measurement. The yields $N_{\text{SM}}^{\ggS}$, $N_{\text{SM}}^{gg\rightarrow 4\ell \text{(box)}}$, and $N_{\text{SM}}^{\ggZZ}$ are those predicted by the Standard Model for only the off-shell Higgs production process, only the box diagram, and the total $\ggB$ contribution including interference, respectively, and are set to the best available prediction as discussed in Section~\ref{sec:MC}. They are allowed to float within the associated theoretical uncertainties discussed in Section~\ref{sec:uncertainties}.
The observed 95$\%$ CL upper limit on the signal strength obtained in this way is 6.5. This agrees with the expected 95$\%$ CL upper limit of 5.4 within the range of [4.2,\,7.2] for $\pm 1 \sigma$ uncertainty.
This extraction demonstrates the degree to which an interpretation of measured cross-sections can approach the precision of dedicated measurements performed at detector level. The result can be compared to the upper limit of 4.5 obtained by the dedicated detector-level measurement~\cite{HIGG-2017-06} in the $4\ell$ final state using the same dataset and the same model.
The sensitivity of this interpretation is slightly lower in comparison, due to the restrictions the unfolding procedure imposes on the binning of observables, the $D_\textrm{ME}$ discriminant in particular.

\subsection*{Constraint on modified Higgs boson couplings}
\begin{figure}[hp]
        \centering
        {\includegraphics[width = 0.72 \textwidth]{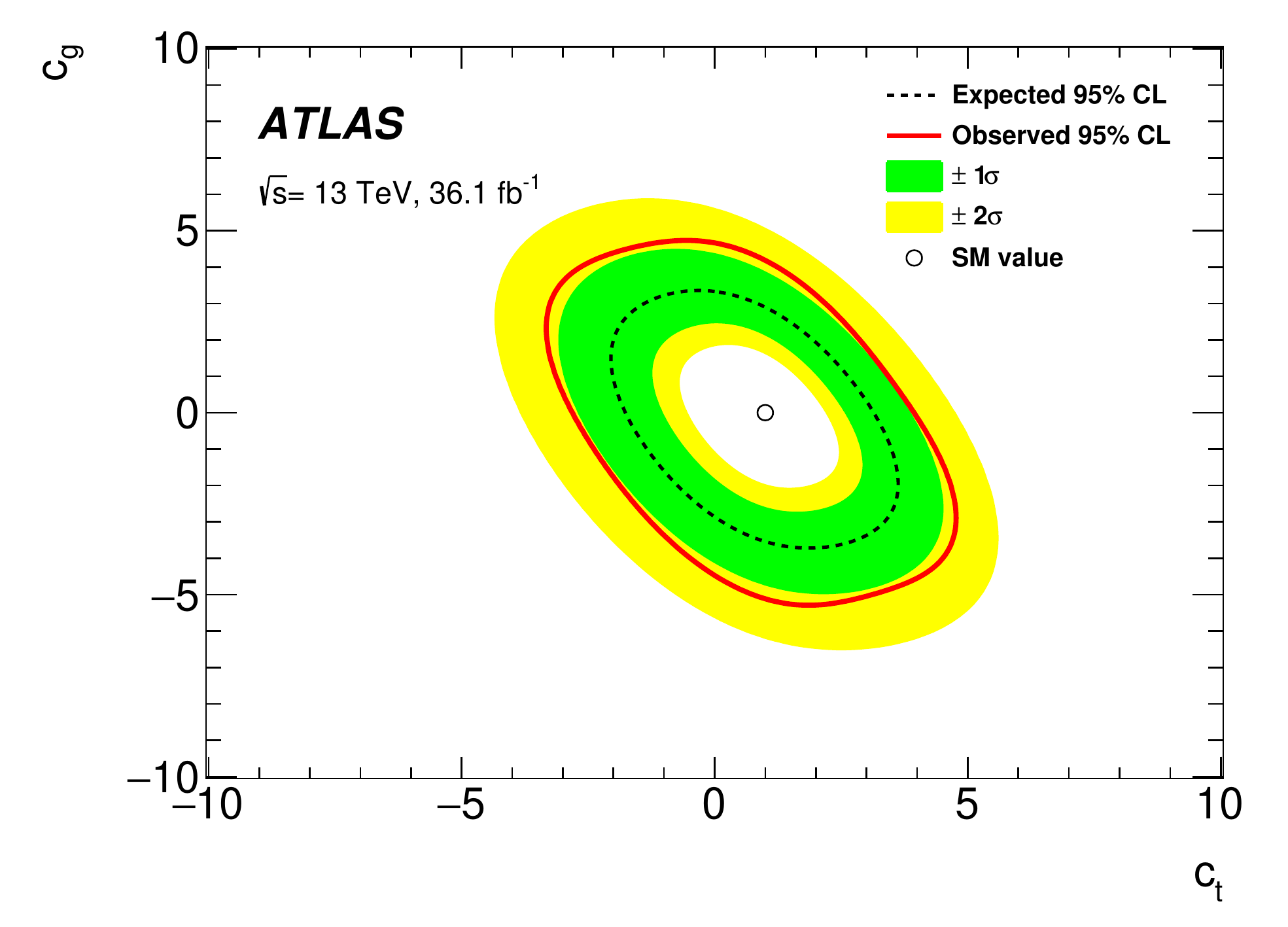}}\\
        \caption{Observed (solid) and expected (dashed) exclusion limits at $95\%$ CL in the $c_{g}$ versus $c_{t}$ plane for modified $t\bar{t}H$ and $ggH$ couplings. The uncertainties in the expected limit corresponding to one and two standard deviations are displayed as green and yellow bands respectively. The hollow circle denotes the tree-level SM values of the parameters: $c_{g}=0$ and $c_{t}=1$. \label{fig:freqEFT} }
\end{figure}

Finally, the detector-corrected four-lepton mass distribution is used to constrain possible BSM modifications of the couplings of the Higgs boson to top quarks ($c_{t}$) and gluons ($c_{g}$, zero in the SM)~\cite{EFT}. 
On-shell rates for Higgs production via gluon--gluon fusion are only sensitive to $|c_{t} + c_{g}|^{2}$, but measurements at higher mass ($>180$~\GeV) can be used to probe these parameters independently, as the partonic centre-of-mass energy of the process becomes larger than the top-quark mass.
This provides an interesting test of the off-shell behaviour beyond dedicated measurements based on the rare $t\bar{t}H$ production mode~\cite{HIGG-2018-13}. 
Again, the yield from $\qqZZ$ is set to the Standard Model prediction and allowed to float within the associated theoretical uncertainties, while the yield from $\ggZZ$ is parameterised as a function of $c_{t}$ and $c_{g}$ using the procedure described in Ref.~\cite{EFT}. 
The observed and expected 95$\%$ CL exclusion contours obtained using the CL$_\text{s}$ method~\cite{CLsMethod} are shown in Figure~\ref{fig:freqEFT}, and the expected limit has green and yellow bands indicating uncertainties of $\pm 1 \sigma$ and $\pm 2 \sigma$.
The parameter space which lies outside of the observed contour is excluded at 95$\%$ CL.
     
Exclusion limits were also explored for a model of anomalous triple gauge couplings considered in a dedicated search region of the ATLAS on-shell $ZZ\to 4\ell$ 
measurement~\cite{STDM-2016-15}. 
Here, it was found that the present detector-corrected analysis is far less sensitive.
This is a general feature of cross-section measurement reinterpretations in terms of models with effects that appear in the very poorly populated tails of distributions: the statistical requirements of unfolding mean that bins will need to be wide in these regions, and therefore sensitivity will be decreased.
\FloatBarrier

\section{Conclusion}
\label{sec:conclusion}
The four-lepton mass distribution has been measured using 36.1~fb$^{-1}$ of proton--proton collision data at a centre-of-mass energy of $\sqrt{s}=$13~\TeV\ recorded with the ATLAS detector at the LHC.
The measurement is made differentially in the invariant mass $\mFourL$ of the four-lepton system, and double-differentially as a function of $\mFourL$ versus
the transverse momentum of the four-lepton system, the rapidity of the system, the matrix-element discriminant $D_\textrm{ME}$ designed to isolate off-shell Higgs boson contributions, and the final state lepton flavour channel.

The measurements are consistent with the predictions of the SM. 
All measurements made are readily reinterpretable in terms of improved SM calculations or additional BSM scenarios. 
A range of example interpretations are presented to demonstrate and explore this potential. 
The signal strength of the gluon--gluon fusion production process is measured to be  $\mu_{gg} = 1.3 \pm 0.5$ compared to an expected value of $1.0 \pm 0.4$.
A value for the $Z \rightarrow 4\ell$ branching fraction of [4.70$\pm$0.32(stat)$\pm$0.25(syst)]$\times 10^{-6}$ is obtained, consistent with existing measurements and exceeding the precision of previous ATLAS results. 
An upper limit on the signal strength for the off-shell Higgs production process of $\mu_H^{\text{OS}} < 6.5$ is obtained at 95\% CL. 
Finally, limits on anomalous couplings of the Higgs boson to gluons and top quarks are derived.

\section*{Acknowledgements}

We thank CERN for the very successful operation of the LHC, as well as the
support staff from our institutions without whom ATLAS could not be
operated efficiently.

We acknowledge the support of ANPCyT, Argentina; YerPhI, Armenia; ARC, Australia; BMWFW and FWF, Austria; ANAS, Azerbaijan; SSTC, Belarus; CNPq and FAPESP, Brazil; NSERC, NRC and CFI, Canada; CERN; CONICYT, Chile; CAS, MOST and NSFC, China; COLCIENCIAS, Colombia; MSMT CR, MPO CR and VSC CR, Czech Republic; DNRF and DNSRC, Denmark; IN2P3-CNRS, CEA-DRF/IRFU, France; SRNSFG, Georgia; BMBF, HGF, and MPG, Germany; GSRT, Greece; RGC, Hong Kong SAR, China; ISF and Benoziyo Center, Israel; INFN, Italy; MEXT and JSPS, Japan; CNRST, Morocco; NWO, Netherlands; RCN, Norway; MNiSW and NCN, Poland; FCT, Portugal; MNE/IFA, Romania; MES of Russia and NRC KI, Russian Federation; JINR; MESTD, Serbia; MSSR, Slovakia; ARRS and MIZ\v{S}, Slovenia; DST/NRF, South Africa; MINECO, Spain; SRC and Wallenberg Foundation, Sweden; SERI, SNSF and Cantons of Bern and Geneva, Switzerland; MOST, Taiwan; TAEK, Turkey; STFC, United Kingdom; DOE and NSF, United States of America. In addition, individual groups and members have received support from BCKDF, CANARIE, CRC and Compute Canada, Canada; COST, ERC, ERDF, Horizon 2020, and Marie Sk{\l}odowska-Curie Actions, European Union; Investissements d' Avenir Labex and Idex, ANR, France; DFG and AvH Foundation, Germany; Herakleitos, Thales and Aristeia programmes co-financed by EU-ESF and the Greek NSRF, Greece; BSF-NSF and GIF, Israel; CERCA Programme Generalitat de Catalunya, Spain; The Royal Society and Leverhulme Trust, United Kingdom. 

The crucial computing support from all WLCG partners is acknowledged gratefully, in particular from CERN, the ATLAS Tier-1 facilities at TRIUMF (Canada), NDGF (Denmark, Norway, Sweden), CC-IN2P3 (France), KIT/GridKA (Germany), INFN-CNAF (Italy), NL-T1 (Netherlands), PIC (Spain), ASGC (Taiwan), RAL (UK) and BNL (USA), the Tier-2 facilities worldwide and large non-WLCG resource providers. Major contributors of computing resources are listed in Ref.~\cite{ATL-GEN-PUB-2016-002}.

\clearpage
\printbibliography
\clearpage
 
\begin{flushleft}
{\Large The ATLAS Collaboration}

\bigskip

M.~Aaboud$^\textrm{\scriptsize 34d}$,    
G.~Aad$^\textrm{\scriptsize 99}$,    
B.~Abbott$^\textrm{\scriptsize 125}$,    
O.~Abdinov$^\textrm{\scriptsize 13,*}$,    
B.~Abeloos$^\textrm{\scriptsize 129}$,    
D.K.~Abhayasinghe$^\textrm{\scriptsize 91}$,    
S.H.~Abidi$^\textrm{\scriptsize 164}$,    
O.S.~AbouZeid$^\textrm{\scriptsize 39}$,    
N.L.~Abraham$^\textrm{\scriptsize 153}$,    
H.~Abramowicz$^\textrm{\scriptsize 158}$
% ...
\bigskip
\\

% ...
$^{13}$Institute of Physics, Azerbaijan Academy of Sciences, Baku; Azerbaijan.\\
% ...
$^{34}$$^{(a)}$Facult\'e des Sciences Ain Chock, R\'eseau Universitaire de Physique des Hautes Energies - Universit\'e Hassan II, Casablanca;$^{(b)}$Centre National de l'Energie des Sciences Techniques Nucleaires (CNESTEN), Rabat;$^{(c)}$Facult\'e des Sciences Semlalia, Universit\'e Cadi Ayyad, LPHEA-Marrakech;$^{(d)}$Facult\'e des Sciences, Universit\'e Mohamed Premier and LPTPM, Oujda;$^{(e)}$Facult\'e des sciences, Universit\'e Mohammed V, Rabat; Morocco.\\
% ...
$^{39}$Niels Bohr Institute, University of Copenhagen, Copenhagen; Denmark.\\
% ...
$^{91}$Department of Physics, Royal Holloway University of London, Egham; United Kingdom.\\
% ...
$^{99}$CPPM, Aix-Marseille Universit\'e, CNRS/IN2P3, Marseille; France.\\
% ...
$^{125}$Homer L. Dodge Department of Physics and Astronomy, University of Oklahoma, Norman OK; United States of America.\\
% ...
$^{129}$LAL, Universit\'e Paris-Sud, CNRS/IN2P3, Universit\'e Paris-Saclay, Orsay; France.\\
% ...
$^{153}$Department of Physics and Astronomy, University of Sussex, Brighton; United Kingdom.\\
% ...
$^{158}$Raymond and Beverly Sackler School of Physics and Astronomy, Tel Aviv University, Tel Aviv; Israel.\\
% ...
$^{164}$Department of Physics, University of Toronto, Toronto ON; Canada.\\
% ...
$^{a}$ Also at Borough of Manhattan Community College, City University of New York, NY; United States of America.\\
$^{b}$ Also at California State University, East Bay; United States of America.\\
$^{c}$ Also at Centre for High Performance Computing, CSIR Campus, Rosebank, Cape Town; South Africa.\\
$^{d}$ Also at CERN, Geneva; Switzerland.\\
$^{e}$ Also at CPPM, Aix-Marseille Universit\'e, CNRS/IN2P3, Marseille; France.\\
% ...
$^{*}$ Deceased

\end{flushleft}


\end{document}